\documentclass[10pt,conference]{IEEEtran}

\usepackage{cite}
\usepackage{wrapfig}

\usepackage{graphicx}
\usepackage{dashrule}

\usepackage[linesnumbered,ruled,vlined]{algorithm2e}%[ruled,vlined]{
\usepackage{algpseudocode}
\usepackage{amsmath,amsthm,amssymb}
\usepackage{scalerel}
\usepackage{multirow}
\usepackage{xspace}
\usepackage{hhline}
\usepackage{comment}
\usepackage[caption=false]{subfig}
\usepackage{todonotes}

\usepackage{url}
\DeclareFixedFont{\ttb}{T1}{txtt}{bx}{n}{10} % for bold
\DeclareFixedFont{\ttm}{T1}{txtt}{m}{n}{10}  % for normal

\usepackage{color}
\definecolor{deepblue}{rgb}{0,0,0.5}
\definecolor{deepred}{rgb}{0.6,0,0}
\definecolor{deepgreen}{rgb}{0,0.5,0}

\usepackage{comment}

\usepackage{listings}

\newcommand\pythonstyle{\lstset{
		language=Python,
		basicstyle=\ttm,
		otherkeywords={self},             % Add keywords here
		keywordstyle=\ttb\color{deepblue},
		emph={MyClass,__init__},          % Custom highlighting
		emphstyle=\ttb\color{deepred},    % Custom highlighting style
		stringstyle=\color{deepgreen},
		frame=tb,                         % Any extra options here
		showstringspaces=false,            % 
		numbers=left
}}

\lstnewenvironment{python}[1][]
{
	\pythonstyle
	\lstset{#1}
}
{}

\usepackage{bm}
\usepackage{mathrsfs}
\newcommand{\cordequivvar}[1]{\scaleto{ \stackrel{\smash{\scaleto{\rightarrow}{#1}}}{\sim}} {7pt} }

\newcommand{\cordequiv}{ {\cordequivvar{2.5pt}} }

\newcommand{\cordequivsub}{ {\scaleto{ \stackrel{\smash{\scaleto{\rightarrow}{2pt}}}{\sim}} {7pt}}  }

\newcommand{\mt}[1]{\text{\tt{#1}}}

\newcommand{\swat}{SWaT\xspace}

\newtheorem{definition}{Definition}
\newtheorem{proposition}{Proposition}
\newtheorem{theorem}{Theorem}

\begin{document}

\title{Finding Causally Different Tests for an\\ Industrial Control System}

\author{\IEEEauthorblockN{Christopher M. Poskitt\IEEEauthorrefmark{1},
Yuqi Chen\IEEEauthorrefmark{2}\IEEEauthorrefmark{3},
Jun Sun\IEEEauthorrefmark{1}, and
Yu Jiang\IEEEauthorrefmark{4}}
\IEEEauthorblockA{\IEEEauthorrefmark{1}Singapore Management University, Singapore}
\IEEEauthorblockA{\IEEEauthorrefmark{2}ShanghaiTech University, China}
\IEEEauthorblockA{\IEEEauthorrefmark{3}Shanghai Engineering Research Center of Energy Efficient and Custom AI IC, China}
\IEEEauthorblockA{\IEEEauthorrefmark{4}Tsinghua University, China}}

% author names and affiliations
% use a multiple column layout for up to three different
% affiliations

\maketitle

\begin{abstract}
    Industrial control systems (ICSs) are types of cyber-physical systems in which programs, written in languages such as ladder logic or structured text, control industrial processes through sensing and actuating.
    Given the use of ICSs in critical infrastructure, it is important to test their resilience against manipulations of sensor/actuator inputs.
    Unfortunately, existing methods fail to test them comprehensively, as they typically focus on finding the simplest-to-craft manipulations for a testing goal, and are also unable to determine when a test is simply a minor permutation of another, i.e.~based on the same \emph{causal events}.
    In this work, we propose a guided fuzzing approach for finding `meaningfully different' tests for an ICS via a general formalisation of sensor/actuator-manipulation strategies.
    Our algorithm identifies the causal events in a test, generalises them to an equivalence class, and then updates the fuzzing strategy so as to find new tests that are \emph{causally different} from those already identified.
    An evaluation of our approach on a real-world water treatment system shows that it is able to find 106\% more causally different tests than the most comparable fuzzer.
    While we focus on diversifying the test suite of an ICS, our formalisation may be useful for other fuzzers that intercept communication channels.
\end{abstract}


\section{Introduction}\label{sec:introduction}

Industrial control systems (ICSs) are types of cyber-physical systems (CPSs) consisting of programs, written in languages such as ladder logic or structured text, that control industrial processes through sensing and actuating.
While these programs (the cyber part) are often simple when viewed in isolation, the system as a whole is more complex than the sum of its parts due to the integration of multiple physical processes.
Given this high level of complexity, as well their ubiquity in domains such as critical infrastructure, it is important that ICSs are systematically tested and analysed against a diversity of sensor/actuator manipulations, so as to ensure that their various defences (e.g.~anomaly detectors~\cite{Inoue-et_al17a,Aoudi-et_al18a,Kravchik-Shabtai18a,Lin-et_al18a,Carrasco-Wu19a,Adepu-et_al20a}, digital fingerprinters~\cite{Ahmed-et_al20a,Gu-et_al18a,Kneib-Huth18a}, invariant checkers~\cite{Adepu-Mathur18b,Chen-Poskitt-Sun18a,Choi-et_al18a,Giraldo-et_al18a,Yoong-et_al21a}) react appropriately to potential attacks and failures.
Unfortunately, this is challenging when benchmarks do not exist for the system under test, as it takes significant time and expertise to construct quality test suites for ICSs from scratch.

An alternative approach, in the absence of benchmarks, is to assess ICS defences against tools such as `CPS fuzzers'~\cite{Chen-Poskitt-et_al19a,Chen-Xuan-Poskitt-et_al20a,Wijaya-Aniche-Mathur20a,Kim-et_al21a}, which automatically search for potential tests, i.e.~manipulations of different network inputs that drive the system towards a targeted unsafe physical state (e.g.~a dangerously high tank level in a water treatment system).
The fuzzers proposed in our previous work~\cite{Chen-Poskitt-et_al19a,Chen-Xuan-Poskitt-et_al20a}, for example, find tests based on the manipulation of high-level actuator commands or low-level bit flipping in network packets.
In both cases, the search is guided by machine learning~(ML) models that predict the effects of potential manipulations.
Experimental evaluations of these fuzzers show they can generate test suites that cover a similar range of testing goals to those of an established benchmark~\cite{Goh-et_al16a} for a water treatment testbed.

These fuzzers, unfortunately, share a critical common limitation in that they focus on the \emph{destination rather than the journey}: their underlying search algorithms are designed to find \emph{some} test as quickly as possible, but give little attention to the details of its steps, so long as it achieves the testing goal.
First, the fuzzers do not consider whether all the individual manipulations in a test were actually necessary for it to have succeeded, potentially leading to tests that are easier for countermeasures to detect (since they are not as `subtle' as they could have been).
Second, once a test is identified, the fuzzers have no systematic way of determining whether there are different strategies that bring about the \emph{same} unsafe state (i.e.~same testing goal).
This potentially leads to test suites that lack diversity, in that they are dominated by tests that are `easier' for the search algorithms to find.
This is exactly what we observe on an actual benchmark for a real-world water treatment system~\cite{CPS-Datasets}, which was accumulated over a period of several years with the help of industrial domain experts as well as researchers, multiple hackathons (with teams from academia and industry), and fuzzers.
In particular, it only contains at most one or two tests per stated goal.

Addressing these two problems is harder for ICSs and CPSs in general than it might at first seem.
Identifying the necessity of individual manipulations boils down to an analysis of \emph{causality} (or cause and effect), and this is made more complicated by factors such as sequencing of steps.
For example, the manipulation of a particular actuator might only be causal for a test if it takes place after the manipulation of another.
Furthermore, the problem of finding \emph{different} tests with the same goal (i.e.~reaching the same unsafe physical state) is challenging due to the ambiguity of what it means for two tests to be `different'.
Suppose, for example, that manipulation $x$ followed by $y$ brings about a targeted unsafe state.
If $y$ followed by $x$ brings about the same unsafe state, is this to be considered `different', given that the same manipulations are used?
What about a sustained sequence of $x$ followed by $y$?
Or $x$ followed by $x$ and $y$ together?
While all these tests are different in a strict sense, depending on the context, they may not be different in a \emph{meaningful} way.
Current fuzzers, unfortunately, lack any support for this kind of reasoning.

\begin{figure}[!t]
	\centering
	\includegraphics[width=1\linewidth]{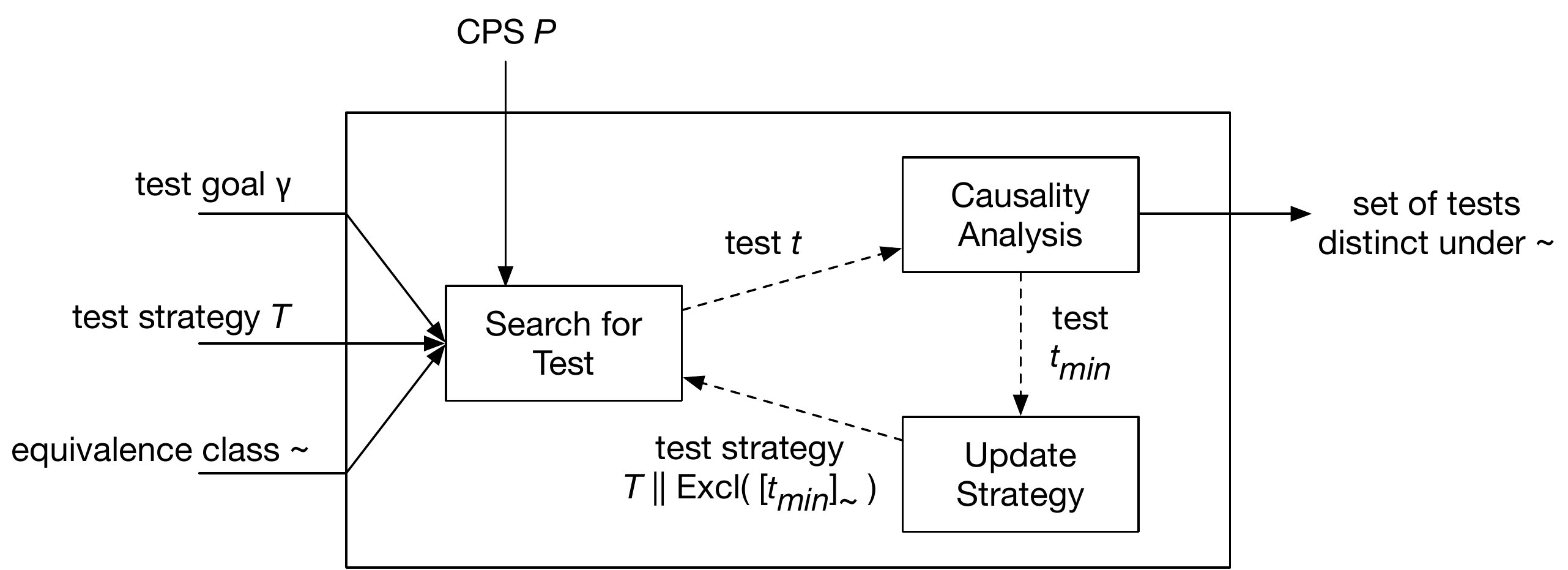}
	\caption{Overview of our approach for finding causally different tests}
	\label{fig:overview_of_approach}
\end{figure}

In this work, we present a comprehensive and practical guided fuzzing approach for finding \emph{meaningfully different} tests, based on a general formalisation of sensor/actuator-manipulation strategies.
Our algorithm identifies the \emph{causal} manipulations in a test, i.e.~without which the test would not have achieved its goal.
It then generalises the test to a user-defined \emph{equivalence class}, which formalises what it means for tests to be the `same' based on common factors such as the set of (causal) manipulations used, or the sequence they were utilised in.
Finally, it updates its fuzzing strategy so as to find new tests that are causally different based on the equivalence classes previously identified.

Figure~\ref{fig:overview_of_approach} gives an overview of this general approach, showing how a test strategy $\mathcal{T}$ is updated to exclude tests equivalent to a previous one $t$ under equivalence class $\sim$.
Our underlying algorithm is based on the idea of modelling test strategies as transition systems labelled with the sensor/actuator values that can be manipulated.
After deriving a test from a strategy, the strategy is updated to guide the search away from tests that simply belong to the same equivalence class, ensuring the next test succeeds for different reasons.
This ability to exclude causally equivalent tests from the search is the key contribution of our formalism that differs it from other test case generation approaches on labelled transition systems~(LTSs) (e.g.~\cite{Tretmans96a}).

We demonstrate the practical viability of our formalisation and approach by implementing a \emph{causal fuzzer} for the Secure Water Treatment (SWaT) testbed~\cite{Mathur-Tippenhauer16a}, a complex multi-stage water purification system based on a real-world plant. First, our experiments show that our fuzzer is able to derive tests for 16 different goals, matching the coverage of previous fuzzers that were applied to this ICS.
Second, but most importantly, we find 33 \emph{causally different} tests for achieving those 16 goals, i.e.~106\% more causally different tests than before.
Furthermore, we found that using our causal equivalence classes allowed our fuzzer to reduce the test suite by 4 orders of magnitude while still covering all relevant causal events.

Overall, this paper makes the following contributions:

\begin{itemize}
    \item A general formalisation of test strategies based on LTSs and traces of sensor/actuator manipulations;
    \item Three equivalence classes for characterising what it means for two tests to be the `same';
    \item A linear-time over-approximation algorithm for identifying the causal events in a test;
    \item A goal-driven test generation strategy that combines fuzzing with our LTSs;
    \item An implementation of the approach for a water treatment system that found 33 causally different tests.
\end{itemize}

While our focus is on diversifying the test suite of an ICS, it is possible that our formalisation could be applied to any kind of software/system in which the execution trace can be influenced by intercepting some communication channel.

\section{Case Study: SWaT Testbed}
\label{sec:background}  \label{sec:overview}

\begin{figure*}[!t]
	\centering
	\includegraphics[width=0.85\linewidth]{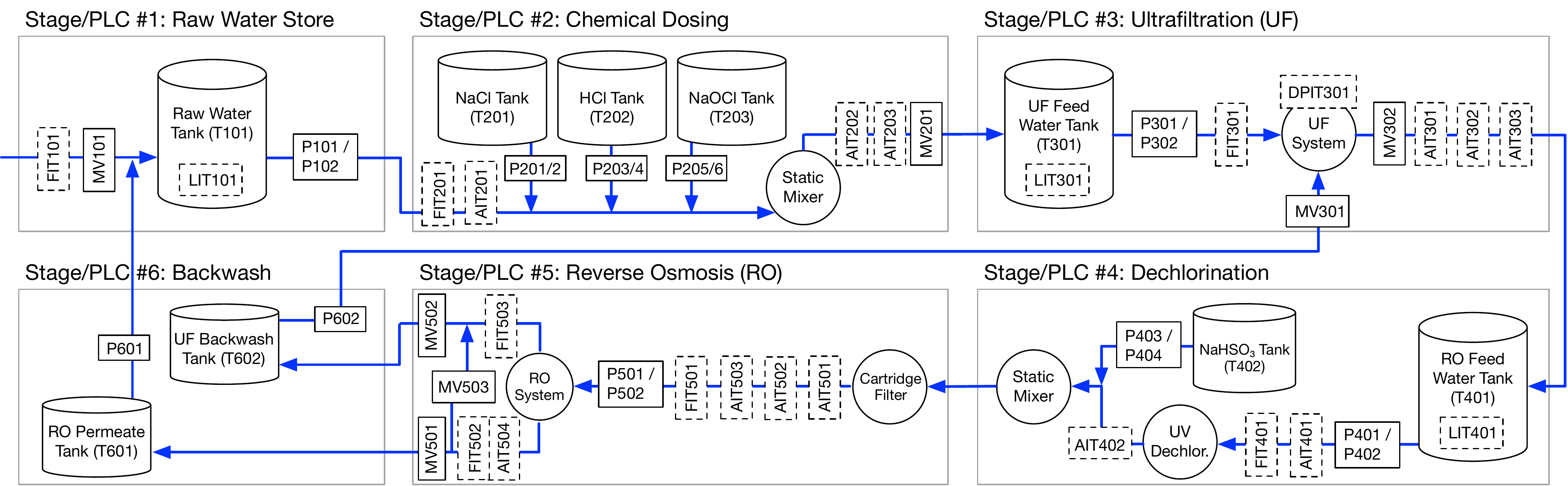}
	\caption{Overview of the six sub-stages of SWaT (blue arrows indicate water flow; dashed/solid rectangles indicate sensors/actuators)}
	\label{fig:swat_overview}
\end{figure*}

We use a complex real-world ICS as our running example: the Secure Water Treatment~(SWaT) testbed~\cite{SWaT-Reference,Mathur-Tippenhauer16a}.
SWaT is a scaled-down version of a working water purification plant, intended to support research on securing and defending critical infrastructure.
SWaT consists of a modern six-stage process that produces up to five gallons of safe drinking water per minute.
Each stage focuses on a particular chemical process (e.g.~dosing, ultrafiltration, or dechlorination) and is controlled by a dedicated Allen-Bradley ControlLogix Programmable Logic Controller~(PLC).
These PLCs repeatedly cycle through their programs, computing the appropriate \emph{commands} to send to actuators based on the latest sensor readings received.
The testbed consists of 36 sensors in total, including water flow indicator transmitters (FITs), tank level indicator transmitters (LITs), and chemical analyser indicator transmitters (AITs).
Among the 30 actuators are motorised valves (MVs) for controlling the inflow of water to tanks, and pumps (Ps) for pumping the water out.
Figure~\ref{fig:swat_overview} provides an overview of the six stages, as well as the main sensors and actuators involved.

The network of the SWaT testbed is organised into a layered hierarchy compliant with the ISA99 standard, providing different levels of segmentation and traffic control. The `upper' layers of the hierarchy, Levels 3 and 2, respectively handle operation management (e.g.~the historian server that logs data) and supervisory control (e.g.~touch panel, engineering workstation). Level 1 is a star network connecting the PLCs, and implements the Common Industrial Protocol~(CIP) over EtherNet/IP. Finally, the `lowest' layer of the hierarchy is Level 0, which consists of ring networks %(EtherNet/IP over UDP) 
that connect individual PLCs to their relevant sensors and actuators.

Each sensor in SWaT is associated with a manufacturer-defined range of \emph{safe} values, which they are expected to remain within at all times during normal operation. If a sensor reports a (true) reading outside of this range, we say the physical state of the system has become \emph{unsafe}. If the tank level indicator transmitter LIT101 in stage one, for example, reports a reading below 250mm, then the physical state has become unsafe due to the risk of an underflow scenario (which can damage the pumps). Similarly, a pressure reading outside of the safe range indicates that a pipe is at risk of bursting.

SWaT is associated with a number of additional resources beyond the physical testbed itself. First, it has an extensive dataset~\cite{Goh-et_al16a,CPS-Datasets}, consisting of the sensor readings and actuator states observed over seven days of continuous normal operation, and four days of various benchmarked attack scenarios. Furthermore, SWaT has a simulator~\cite{Supplementary-Material} that can be used for offline analyses. Built in Python, it faithfully simulates the control logic of the PLCs, and executes models of some key physical processes (e.g.~water flow). As it was cross-validated against real data from the dataset, the simulator can be used as a way to initially evaluate various over- and underflow attacks without wasting the resources of the physical testbed.

Across the six sub-processes of SWaT, there are several different unsafe physical states as measured by the sensors.
In Stage 3, for example, if the differential pressure transmitter DPIT301 detects a pressure level less than 0.1 bar, the ultrafiltration process is in an unsafe state because the subsequent backwash process will become stuck.
A CPS fuzzer that aims to get DPIT301 below this threshold as soon as possible will typically find the simplest way to do so, i.e.~by manipulating at least the motorised valve MV302.
However, this is not the only way that this unsafe state can be brought about: one can also switch off two pumps (P301 and P302) or open MV301, MV303, and switch on P602.
Finding these alternative tests automatically is difficult without a causal understanding of why the tests succeed, or a way to determine that a test is not simply a permutation of an `equivalent' one.

\section{Systems, Strategies, and Tests}
\label{sec:sys_strat_att}

In order to solve the problem of finding causally different tests for ICSs such as SWaT, we need a formalisation that allows fuzzers to reason precisely about tests, manipulation strategies, and causal events. We thus present an intuitive formal definition of systems, define strategies as LTSs, and formalise tests as traces of sensor/actuator manipulations.

Our formalism aims for generality, modelling the key characteristics of CPSs (e.g.~control states, physical states, and sensing/actuating components) in a way that abstracts away from specific ICS implementation details (e.g.~ladder logic). We remark that while generality is our intention, our current evaluation and case study focus on an ICS representative of the water domain. Assessing the utility of this formalisation for other kinds of CPSs is important future work.

\subsection{Systems}

Intuitively, a CPS consists of software interacting with one or more physical processes via components connected over a network.
In this work, we assume there to be two types of components: sensors, which read continuous data from the physical state (e.g.~temperature, pressure, flow), and actuators, which are mechanised devices for controlling the process (e.g.~motorised pumps or valves).

\begin{definition}[Component; sensor; actuator]\rm
	A \emph{component}~$c$ is a device for interacting with a physical process; it has an internal state derived from an associated \emph{domain} of values $D_c$.
	A \emph{sensor}~$s$ is a component with domain $D_s \subseteq \mathbb{R}$, i.e.~modelling real-valued readings from a process.
	An \emph{actuator}~$a$ is a component with a discrete finite domain $D_a$.
	\qed
\end{definition}

In this paper, if an actuator is a pump, we assume the domain $\{\mathtt{on}, \mathtt{off}\}$, and if it is a motorised valve, we assume the domain $\{\mathtt{open}, \mathtt{closed}\}$.
Actuators with `partial' positions (e.g.~$25\%$ open) would require an appropriate discretisation.% to be chosen.

For a given set of sensors or actuators, we refer to their current states as their readings or configurations respectively.

\begin{definition}[Readings; configurations]\rm
	Given a set of sensors (resp.~actuators) $C$, the \emph{readings} (resp.~\emph{configurations}) of $C$ are denoted as a set of pairs $\overline{C} \subseteq \{ \langle c, v \rangle \mid c \in C, v \in D_c  \}$ containing exactly one pair per component, i.e.~if $\langle c, v_1\rangle \in \overline{{C}}$ and $\langle c, v_2\rangle \in \overline{{C}}$ then $v_1 = v_2$. We let $\mathbb{C}$ denote the set of \emph{all possible} readings/configurations $\overline{{C}}$.
	\qed
\end{definition}

Our definition of a CPS elaborates on how control, physical, and component states all relate to each other.
Note that the transition functions can be left implicit if the targeted system is available to be executed (as is the case for SWaT; Section~\ref{sec:implementation}).

\begin{definition}[Cyber-physical system]\rm
	A \emph{cyber-physical system (CPS)} $\mathcal{P}$ is a tuple of the form $(Q_\mathcal{P}, X_\mathcal{P}, S_\mathcal{P}, A_\mathcal{P}, \delta_\mathcal{P}, d_\mathcal{P}, \theta_\mathcal{P}, \tau_\mathcal{P})$, where $Q_\mathcal{P}$ is a set of control states, $X_\mathcal{P}$ is a set of states of the physical process, $S_\mathcal{P}$ is a set of sensors, $A_\mathcal{P}$ is a set of actuators, $\delta_\mathcal{P}\!: Q_\mathcal{P} \times \mathbb{S}_\mathcal{P} \rightarrow Q_\mathcal{P} \times \mathbb{A}_\mathcal{P}$ is the logic of the controllers, $d_\mathcal{P}\!: X_\mathcal{P} \times \mathbb{A}_\mathcal{P} \rightarrow X_\mathcal{P}$ is a function characterising how the physical process evolves after a fixed time interval $\tau_\mathcal{P}$, and $\theta_\mathcal{P}\!: X_\mathcal{P} \rightarrow \mathbb{S}_\mathcal{P}$ is an observation function describing how sensor readings are extracted from the state of the physical process.
	\qed
\end{definition}

Given a CPS $\mathcal{P}$, a \emph{run} of the system is a sequence of the form $(q_0, x_0) \rightarrow (q_1, x_1) \rightarrow \cdots$ such that each $q_i\in Q_\mathcal{P}$, $x_i\in X_\mathcal{P}$, and for every step $(q_i, x_i) \rightarrow (q_{i+1}, x_{i+1})$, $\delta_\mathcal{P}(q_i, \theta_\mathcal{P}(x_i)) = (q_{i+1}, \overline{{A}}_\mathcal{P})$ with $\overline{{A}}_\mathcal{P} \in \mathbb{A}_\mathcal{P}$ and $d_\mathcal{P}(x_i, \overline{{A}}_\mathcal{P}) = x_{i+1}$.

\subsection{Test Strategies}
CPS testing involves manipulating sensor readings and actuator states according to a defined strategy.
%CPS attackers are entities that manipulate sensor readings and actuator states according to a defined strategy.
Our definition consists of two parts.
First, we define \emph{capabilities}, which are the particular manipulations that are possible.
Second, we define \emph{strategies}, which are transition systems expressing how the capabilities will be utilised, and the conditions of the system that must be true for them to be able to do so.
We can specify a range of test strategies, including optimistic ones that simply `try out' various capabilities, or more sophisticated ones that conduct manipulations based on the system state.
%only conduct attacks depending on the system state.

Capabilities are denoted as pairs of components and values (we use square brackets to distinguish them from readings).
A pair $[ s, v ]$ for sensor $s$ and value $v \in D_s$ indicates that the tester is capable of spoofing the reading reported by $s$ as $v$, regardless of what the actual reading is.
Analogously, a pair $[ a, v ]$ for actuator $a$ and value $v \in D_a$ indicates that the tester is capable of forcing actuator $a$ into configuration $v$, regardless of the commands that should be being issued to the actuator at a given moment of the system's execution.

\begin{definition}[Capability]\rm
	Let $\mathcal{P}$ be a CPS. A \emph{capability over $\mathcal{P}$} is a pair $[ c,v ]$, where $c$ is a component in $S_\mathcal{P} \cup A_\mathcal{P}$ and $v \in D_c$.
	\qed
\end{definition}

A set of capabilities may be infinite if the tester can manipulate sensor values over a continuous domain.
In such cases, we use the notation $[ s, R ]$ for sensor $s$ and $R \subseteq D_s$ to represent all capabilities $[ s, v ]$ for $v \in R$.
For example, $[ s, \mathbb{N}_0 ]$ would represent $[ s, 0 ], [ s,1 ], [ s,2 ]$, and so on.

As an example, the set of capabilities $\{ [ \mathtt{P_{101}}, \mathtt{on} ],$ $[ \mathtt{MV_{101}}, \mathtt{close} ] \}$ over SWaT expresses the ability to override the configurations of two actuators in stage one.
In particular, the capabilities respectively express manipulations that cause pump $\mathtt{P_{101}}$ to switch on and motorised valve $\mathtt{MV_{101}}$ to close.
Note that our model abstracts away from how these capabilities are realised, specifying only that they \emph{can} be.

Test \emph{strategies} are defined as LTSs with labels of the form $\gamma \vdash \varphi$.
Here, $\gamma$ is a \emph{sensor condition} that must be true for the transition to fire, whereas $\varphi$ is a \emph{capability condition} that constrains the capabilities that it is allowed to utilise.

\begin{definition}[Sensor condition]\rm
	Let $\mathcal{P}$ denote a CPS. A \emph{sensor condition} over $\mathcal{P}$ is a Boolean formula of simple linear inequalities over the sensors $S_\mathcal{P}$. The set of all possible sensor conditions over $\mathcal{P}$ is denoted by $\mathrm{Cond}_{\mathcal{P}}$. Given a condition $\gamma \in \mathrm{Cond}_{\mathcal{P}}$ and physical state $x \in X_\mathcal{P}$, the \emph{valuation} of the condition, $\gamma^{x} \in \mathbb{B}$, is obtained by evaluating the Boolean expression in the standard way, but with each sensor symbol $s$ interpreted as $v$ such that $\langle s, v \rangle \in \theta_\mathcal{P} (x)$.
	\qed
\end{definition}

As an example, the sensor condition $\gamma = \mathtt{LIT_{101}} \geq 250 \wedge \mathtt{LIT_{101}} \leq 1100$ over \swat expresses that the level of tank one is between 250mm and 1100mm, i.e.~within its safe range. Given a physical state $x$, the valuation of $\gamma^x$ is true if $250 \leq v \leq 1100$ for $\langle \mathtt{LIT_{101}}, v \rangle \in \theta_\mathtt{SWaT}(x)$, and false otherwise.

Capability conditions are simple Boolean expressions over capability sets.
We use the symbol `$\_$' to represent the capabilities that are utilised when the transition is fired.
Conditions across multiple labels can be related using globally scoped variables from a set Var, e.g.~$\mathtt{X}\in\mathrm{Var}$.
% written in a typewriter font

\begin{definition}[Capability condition]\rm
	Let Cap denote a set of capabilities.
	A \emph{capability condition} over Cap is a Boolean formula of the form true, $\mathrm{Exp}_1 \subseteq \mathrm{Exp}_2$, $\mathrm{Exp}_1 == \mathrm{Exp}_2$, $\varphi_1 \vee \varphi_2$, $\varphi_1 \wedge \varphi_2$, or $\neg \varphi_1$.
	Here, $\mathrm{Exp}_1$, $\mathrm{Exp}_2$ can be expressions over variables in Var, subsets of Cap, or the symbol $\_$, whereas $\varphi_1$, $\varphi_2$ are capability conditions.
	We let $\mathrm{Cond}_\mathrm{Cap}$ denote the set of all possible capability conditions over Cap.
	An \emph{assignment} is a function $\alpha\!: \mathrm{Var} \rightarrow 2^{\mathrm{Cap}}$ mapping variables to subsets of Cap.
	Given a capability condition $\varphi$, a set of capabilities $Y\in 2^\mathrm{Cap}$, and an assignment $\alpha$, the condition's \emph{valuation} $\varphi^{Y,\alpha} \in \mathbb{B}$ is obtained by evaluating the set and Boolean expressions in the usual way, but with $Y$ substituted for $\_$, and variables $\mathtt{X}\in \mathrm{Var}$ interpreted as $\alpha(\mathtt{X})$.
	\qed
\end{definition}

For example, the capability condition $\varphi = [ \mathtt{P_{101}}, \mathtt{on} ] \notin \_$ specifies that any combination of capabilities can be used, so long as $[ \mathtt{P_{101}}, \mathtt{on} ]$ is not included.
Note that we will use $y \in \mathrm{Exp}$, $\mathrm{Exp}_1 \not\subseteq \mathrm{Exp}_2$, and $\mathrm{Exp}_1 \neq \mathrm{Exp}_2$ to respectively abbreviate $\{y\} \subseteq \mathrm{Exp}$, $\neg \mathrm{Exp}_1 \subseteq \mathrm{Exp}_2$, and $\neg (\mathrm{Exp}_1 == \mathrm{Exp}_2)$.
Furthermore, we let the capability set $Y$ on its own abbreviate the capability condition $\_ == Y$.

Finally, we define strategies as LTSs, expressing the possible sequences of capabilities that can be utilised by the tester.

\begin{definition}[Test strategy]\rm
	Let $\mathcal{P}$ denote a CPS and Cap a set of capabilities over $\mathcal{P}$.
	A \emph{test strategy} for $\mathcal{P}$ and Cap is a tuple $\mathcal{T} = (Q_\mathcal{T}, \longrightarrow_\mathcal{T}, i_\mathcal{T})$ where $Q_\mathcal{T}$ is a set of states, $\longrightarrow_\mathcal{T}\ \subseteq Q_\mathcal{T} \times (\mathrm{Cond}_{\mathcal{P}} \times \mathrm{Cond}_{\mathrm{Cap}}) \times Q_\mathcal{T}$ is a transition relation, and $i_\mathcal{T} \in Q_\mathcal{T}$ is the initial state.
	A transition $(r_i, (\gamma, \varphi), r_j) \in\ \longrightarrow_\mathcal{T}$ is visualised as an arrow from $r_i$ to $r_j$ with the label $\gamma \vdash \varphi$ (we may omit $\gamma$ or $\varphi$ when they are true). 
	\qed
\end{definition}

For simplicity, we make some assumptions on the form of strategies.
First, we assume that it is \emph{always} possible to fire at least one transition, regardless of the state of the strategy or the physical process.
(One can always add a `do nothing' transition that uses no capabilities.)
Second, we assume that between any pair of states $r_1, r_2$, there is at most one transition from $r_1$ to $r_2$.
As a consequence, the sequence of transitions fired can be uniquely determined from a sequence of states.
Finally, if a capability set is specified in a transition, then it contains at most one capability per sensor/actuator.

\begin{figure}[!t]
	\centering
	\includegraphics[width=0.9\linewidth]{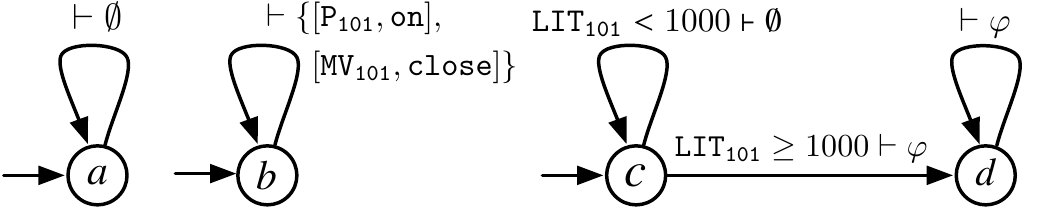}
	\caption{Three examples of test strategies for \swat}
	\label{fig:attack_strategies}
\end{figure}

This definition allows us to capture a variety of test strategies, as shown by the examples in Figure~\ref{fig:attack_strategies}.
The first transition system $(\{a\}, \{ (a, (\mathrm{true}, \_ == \emptyset), a) \}, a)$ implements a \emph{null} strategy.
The sensor condition of the sole transition is vacuously satisfied, but no capabilities are ever utilised.
The second transition system $(\{b\}, \{ (b, (\mathrm{true}, \_==\{ [ \mathtt{P_{101}}, \mathtt{on} ], [ \mathtt{MV_{101}}, \mathtt{close}] \}), b) \},$ $ b)$ implements a \emph{sustained} test strategy.
There is only ever one transition to fire (with a condition that is always satisfied), and it utilises the capability set $\{ [ \mathtt{P_{101}}, \mathtt{on} ],$ $[ \mathtt{MV_{101}}, \mathtt{close} ] \}$ for \swat, i.e.~overriding the configurations of pump $\mathtt{P_{101}}$ (to switch it on) and motorised valve $\mathtt{MV_{101}}$ (to close it), sustaining this for the whole test.

Finally, let $\varphi = [\mathtt{MV_{101}}, \mathtt{open}] \in \_ \wedge \mathtt{X}\ ==\ \_$.
The rightmost transition system in Figure~\ref{fig:attack_strategies} then implements a \emph{conditional} strategy.
For as long as the actual sensor reading of $\mathtt{LIT_{101}}$ remains under 1000mm, no capabilities are used.
As soon as that threshold is crossed during \swat's operation, the transition from $c$ to $d$ is fired, and a set of capabilities is used that satisfies $\varphi$, i.e.~includes at least the opening of motorised valve $\mathtt{MV_{101}}$.
Due to the global variable $\mathtt{X}$ in $\varphi$, that same set of capabilities is used for the rest of the test, i.e.~the looping transition incident to $d$.

\subsection{Tests}
\label{subsec:attacks}

Next, we define \emph{tests}, the specific sequence of manipulations derived from a strategy.
A test can be thought of as a run of the system in which the transitions of a strategy are fired and tracked in parallel.
The key difference is that the capabilities specified in strategies result in \emph{modified observation functions} and \emph{modified actuator commands}, and thus the resulting sequence can contain control and physical states that would not normally be observed in a run of the system.

We begin by defining how the utilisation of capabilities modifies the observation functions and actuator commands of systems.
Given a CPS $\mathcal{P}$ and a capability set $Y$, the \emph{usage of sensor capabilities in $Y$} is reflected by a modified observation function, $\theta_\mathcal{P}^Y$, defined the same as $\theta_\mathcal{P}(x)$ for all $x \in X_\mathcal{P}$, but with elements $\langle s, v \rangle$ replaced by $\langle s, v' \rangle$ if $[s, v'] \in Y$.
Analogously, given a set of actuator configurations $\overline{A}_\mathcal{P} \in \mathbb{A}_\mathcal{P}$, the \emph{usage of actuator capabilities in $Y$} is reflected by a modified set of configurations, $\overline{A}_\mathcal{P}^Y$, obtained from $\overline{A}_\mathcal{P}$ by replacing elements $\langle a, v \rangle$ with $\langle a, v' \rangle$ if $[a, v'] \in Y$.

Intuitively, a test is a sequence of synchronised steps between the control/physical states of the CPS and the states of a strategy.
If a transition in the strategy is labelled with capabilities, they are utilised in the step directly.
If instead it is labelled with $\vdash\varphi$, then \emph{any} combination of capabilities that satisfies $\varphi$ is utilised.

\begin{definition}[Test]\rm
	Given a CPS $\mathcal{P}$, a set of capabilities Cap, and a strategy $\mathcal{T}$, a \emph{test} on $\mathcal{P}$ is a sequence of the form $(q_0, x_0, r_0) \rightarrow_{Y_1} (q_1, x_1, r_1) \rightarrow_{Y_2} \cdots$ where $r_0 = i_\mathcal{T}$ and each $q_i \in Q_\mathcal{P}$, $x_i \in X_\mathcal{P}$, $r_i \in Q_\mathcal{T}$, and $Y_i \in 2^{\mathrm{Cap}}$.
	Furthermore, there is an assignment $\alpha$ such that for every $(q_{i-1}, x_{i-1}, r_{i-1}) \rightarrow_{Y_i} (q_{i}, x_{i}, r_{i})$, there exists a transition $(r_{i-1}, (\gamma_i, \varphi_i), r_{i}) \in\ \longrightarrow_\mathcal{T}$ with $\gamma_i^{x_{i-1}} = \varphi_i^{Y_i,\alpha} = \mathrm{true}$, $\delta_\mathcal{P}(q_{i-1}, \theta^{Y_i}_\mathcal{P}(x_{i-1})) = (q_{i}, \overline{A}_\mathcal{P})$, and $d_\mathcal{P}(x_{i-1}, \overline{A}_\mathcal{P}^{Y_i}) = x_{i}$.
	\qed
\end{definition}

Given a CPS $\mathcal{P}$, a strategy $\mathcal{T}$, and a \emph{testing goal} $\gamma_g \in \mathrm{Cond}_\mathcal{P}$, a \emph{successful test} for $\gamma_g$ is a finite test $(q_0, x_0, i_\mathcal{T}) \rightarrow_{Y_1} \dots \rightarrow_{Y_n} (q_n, x_n, r_n)$ such that $\gamma_g^{x_n}$ is true. When analysing a successful test, it is helpful to be able to reason about the \emph{capability history}, i.e.~the trace of capability sets utilised.
Given a test $t = (q_0, x_0, r_0) \rightarrow_{Y_1} (q_1, x_1, r_1) \rightarrow_{Y_2} \cdots$, the \emph{capability history} of $t$, denoted $\pi_t$, is the corresponding sequence of capability sets $Y_1Y_2 \cdots$.

For a capability history $\pi_t = Y_1 Y_2 \cdots$, we use the notation $\pi_t[k..l]$ with $1 \leq k \leq l$ to denote the \emph{slice} of $\pi_t$ from position $k$ to $l$ inclusive, i.e.~$Y_k Y_{k+1} \cdots Y_{l-1} Y_l$. Furthermore, we define $\mathrm{CSet}(\pi_t) = \bigcup_{i\in\{1,2,\cdots\}} Y_i$ as the cumulative set of all capabilities used in $\pi_t$.

\section{Identifying Equivalence Classes of Tests}
\label{sec:ident_classes}

In this section, we address the problem of identifying classes of equivalent tests and performing online updates of strategies.

First, we propose a number of equivalence classes for expressing different ways that two tests can be related.
In particular, we define classes based on the particular capabilities used and the order in which they are applied.
These specific classes are defined to support our case study (SWaT), but in general, can be defined according to the user's needs.

Second, we show how a strategy can be updated to exclude the possibility of deriving further tests from a given class.
Implementing this step provides testers (e.g.~fuzzers) a mechanism by which subsequent tests can be diversified once a given equivalence class is suitably covered.

\subsection{Equivalence Classes}
\label{sec:equivalence_classes}
We propose three initial equivalence classes that characterise some informal common understandings of what makes two tests the `same'.
In particular, whether two tests both use a given (causal) subset of capabilities, whether they use exactly the same ones, or whether they use them in the same order.

The first of our classes considers two successful tests equivalent if the capabilities they used include a given subset, regardless of the order or context of their application.
This equivalence class is particularly useful for our implemented fuzzer (Section~\ref{sec:approximation}), which is able to approximate the \emph{causal} capabilities of a test, and thus can treat any other test that also uses them as equivalent.

Note that these initial equivalence classes were proposed after examining the existing benchmark of manually constructed SWaT attacks~\cite{Goh-et_al16a} and forming an intuition as to when attacks with the same goal (e.g.~overflow a tank) were considered distinct by the benchmark's designers. For ICSs and CPSs outside of the water domain, the user may need to define different equivalence classes if the specific context requires different criteria to distinguish two tests.

\begin{definition}[Capability-set equivalence]\rm
	Let $t_1,t_2$ denote two successful tests for goal $\gamma_g$, and $Y$ a set of capabilities. We say that $t_1$ and $t_2$ are \emph{$Y$-set equivalent}, denoted $t_1 \simeq_Y t_2$, if $t_1=t_2$, or $Y \subseteq\mathrm{CSet}(\pi_{t_1})$ and $Y \subseteq\mathrm{CSet}(\pi_{t_2})$.
	\qed
\end{definition}

A stronger version of this class considers two successful tests to be equivalent if the tests utilised \emph{exactly the same} capabilities, regardless of when or how they were used.

\begin{definition}[Strong capability-set equivalence]\rm
	Let $t_1,t_2$ denote two successful tests for goal $\gamma_g$. We say that $t_1$ and $t_2$ are \emph{strong capability-set equivalent}, denoted $t_1 \simeq t_2$, if $\mathrm{CSet}(\pi_{t_1}) = \mathrm{CSet}(\pi_{t_2})$.
	\qed
\end{definition}

Our third equivalence class considers two successful tests to be equivalent if their capability histories are equal once duplicate and trailing steps are removed. For example, successful tests with histories $PQQRPP$ and $PPQRRP$ are strong capability-order equivalent, as both contain the same \emph{order} of distinct sets---$PQRP$---when duplicates are removed. Furthermore, if a successful test derives a sequence of distinct sets with $PQRP$ as its prefix (e.g.~$PQRPS$), then that test would be considered equivalent too.

\begin{definition}[Strong capability-order equivalence]\rm
	Let $t_1,t_2$ denote two successful tests for goal $\gamma_g$. We say that $t_1$ and $t_2$ are \emph{capability-order equivalent}, denoted $t_1\!\ \cordequiv\!\ \!\ t_2$, if $\mathrm{COrd}(\pi_{t_1})$ is a prefix of $\mathrm{COrd}(\pi_{t_2})$ (or vice versa).
	Here, we inductively define $\mathrm{COrd}(Y_1Y_2Y_3\cdots) = Y_1\cdot\mathrm{COrd}(Y_kY_{k+1}\cdots)$ where $\cdot$ denotes concatenation, $k$ is chosen such that $Y_1 \neq Y_k$, and for every $k'$ in $1 < k' < k$, $Y_1 = Y_{k'}$.
	For finite capability histories, COrd terminates on the empty sequence.
	\qed
\end{definition} 

As a notational convenience, for a given equivalence class $\sim$, we let $[t]_\sim$ denote the (potentially infinite) set $\{t' \mid t' \sim t \}$.

\subsection{Excluding Classes of Tests}
\label{sec:finding_different_attacks}

Given a strategy $\mathcal{T}$, a successful test $t$, and an equivalence class $\sim$, our goal is to transform $\mathcal{T}$ into a new strategy in which the same tests can be derived \emph{except} for those equivalent to $t$, i.e.~in $[t]_\sim$.
This mechanism allows fuzzers to diversify by `nudging' the search towards tests that are not simply minor permutations of $t$.
Our algorithm involves two steps. First, we construct a strategy $\mathrm{Excl}([t]_\sim)$ that can derive every test excluding those that are equivalent to any in $[t]_\sim$.
Second, we construct a \emph{parallel composition} $\mathcal{T} \parallel \mathrm{Excl}([t]_\sim)$ of the strategies to return a new one that characterises all of the tests possible in $\mathcal{T}$ except for those belonging to $[t]_\sim$.

%We present the construction of $\mathrm{Excl}([t]_\sim)$ for capability-set equivalence (the others are available online~\cite{Double-Blind-Appendix})).
We present the construction of $\mathrm{Excl}([t]_\sim)$ for capability-set equivalence (the others are available in the Appendix).
In the following, we use the notation $L(\mathcal{T})$ to denote the language of all finite capability histories that can be derived from $\mathcal{T}$.

\begin{proposition}[Capability-set equivalence]\label{prop:capability-set_equivalence}
	\emph{Let $t$ denote a successful test on CPS $\mathcal{P}$, and $Y \subseteq \mathrm{CSet}(\pi_t)$ a set of capabilities.
	A strategy $\mathrm{Excl}([t]_{\simeq_Y})$ can be constructed such that for every successful test $t'$ on $\mathcal{P}$:}
	\[ \pi_{t'} \in L(\mathrm{Excl}([t]_{\simeq_Y}))\ \ \text{\emph{if and only if}}\ \ t' \not\in [t]_{\simeq_Y} \]

	\noindent Construction. \emph{Suppose that $Y = \{y_1, \dots, y_n \}$. Then:}
\begin{equation*}
    \footnotesize
	\begin{split}
		\mathrm{Excl}([t]_{\simeq_Y}) =\ &(\{q_0, q_1, \dots, q_n\},\\
		& \{(q_0, (\mathrm{true}, (y_1,\dots,y_n \not\in \_)), q_0)\}\vspace{-0.5in} \\%[-8pt]
		&\ \ \ \bigcup\limits_{i=1}^{n} \bigl\{ (q_0, (\mathrm{true}, y_i \notin \_), q_i), (q_i, (\mathrm{true},y_i \notin \_), q_i) \bigr\},\\
		& q_0)
	\end{split}
\end{equation*} % & \hspace{0.3in}(q_i, (\mathrm{true},\_, y_i \notin \_), q_i) \bigr\}, q_0)

\noindent \emph{where $q_0, q_1, \dots, q_n$ are fresh strategy states.}
\qed
\end{proposition}

\begin{figure}[!t]
	\centering
	\includegraphics[width=0.6\linewidth]{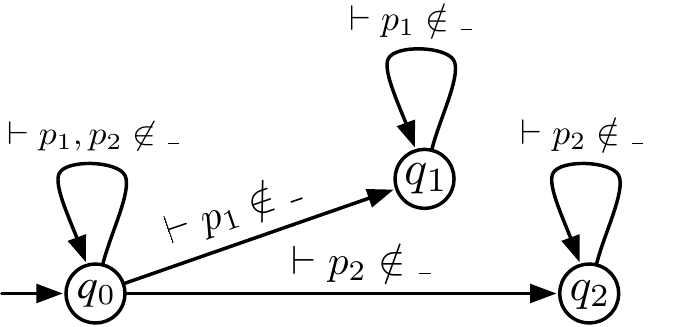}
	\caption{Constructing $\mathrm{Excl}([t]_{\simeq_Y})$ for capability set $Y = \{p_1, p_2\}$}
	\label{fig:construction_capset}
\end{figure}

For example, suppose that $\pi_t = PPPQQQPPP$ is the history of a successful test $t$, where $P = \{p_1\}$ and $Q = \{p_1,p_2\}$.
Let $Y = \{p_1,p_2\}$.
Then $\mathrm{Excl}([t]_{\simeq_Y})$ is as depicted in Figure~\ref{fig:construction_capset}.
Intuitively, the strategy characterises three types of tests: those in which $p_1$ and $p_2$ are never used; those in which $p_1$ is used at least once but $p_2$ never is; and those in which $p_2$ is used at least once but $p_1$ never is.
In particular, the strategy excludes any tests in which \emph{both} $p_1$ and $p_2$ are used at some point (whether together or separately), i.e.~the class of all tests that are $Y$-set equivalent to $t$.

Next, we define a parallel composition operator $\parallel$ for strategies.
In particular, it allows us to construct $\mathcal{T} \parallel \mathrm{Excl}([t]_\sim)$, i.e.~a strategy allowing all of the tests of $\mathcal{T}$ except those equivalent to $t$ under $\sim$.

\begin{definition}[Strategy composition]\rm
	Let $\mathcal{P}$ denote a CPS. Suppose that $\mathcal{T}_1,\mathcal{T}_2$ are strategies for $\mathcal{P}$ with disjoint variables and $Q_{\mathcal{T}_1} \cap Q_{\mathcal{T}_2} = \emptyset$.
	Then the \emph{composition} of the strategies, denoted $\mathcal{T}_1 \parallel \mathcal{T}_2$, is the tuple $(Q_{\mathcal{T}_1} \times Q_{\mathcal{T}_2}, \longrightarrow_{\mathcal{T}_1 \parallel \mathcal{T}_2}, (i_{\mathcal{T}_1}, i_{\mathcal{T}_2}))$ where:
	\begin{equation*}
		\begin{split}
			\longrightarrow_{\mathcal{T}_1 \parallel \mathcal{T}_2}\ =\ &\{ ((q_i,r_k), (\gamma_1 \wedge \gamma_2,\varphi_1 \wedge \varphi_2), (q_j,r_l))\\
			& \mid\ (q_i, (\gamma_1, \varphi_1), q_j) \in\ \longrightarrow_{\mathcal{T}_1} \\
			& \hspace{0.1in}\wedge\ (r_k, (\gamma_2, \varphi_2), r_l) \in\ \longrightarrow_{\mathcal{T}_2} \} \hfill \qed
		\end{split} 	
	\end{equation*}

\end{definition}

\begin{figure}[!t]
	\centering
	\includegraphics[width=0.4\linewidth]{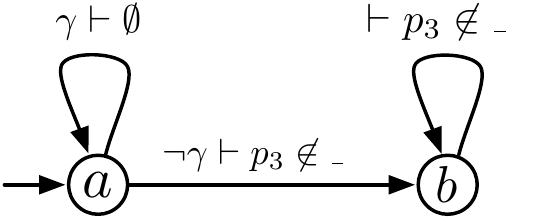}
	\caption{Example test strategy $\mathcal{T}$}
	\label{fig:parallel_example_strategy}
\end{figure}

To illustrate, consider the strategy $\mathcal{T}$ of Figure~\ref{fig:parallel_example_strategy}, in which any capability (except $p_3$) may be used as soon as some sensor condition $\gamma$ no longer holds.
Suppose that $\emptyset\emptyset\emptyset PPQQQQ$ is the history of a successful test $\pi_t \in L(\mathcal{T})$, where $P = \{p_1\}$ and $Q = \{p_1, p_2\}$.
Let $Y = \{ p_1, p_2 \}$.
Suppose that we want to find tests that are \emph{not} $Y$-set equivalent to $t$ (perhaps, for example, because $p_1,p_2$ are causal; see Section~\ref{sec:approximation}).

\begin{figure}[!t]
	\centering
	\includegraphics[width=0.7\linewidth]{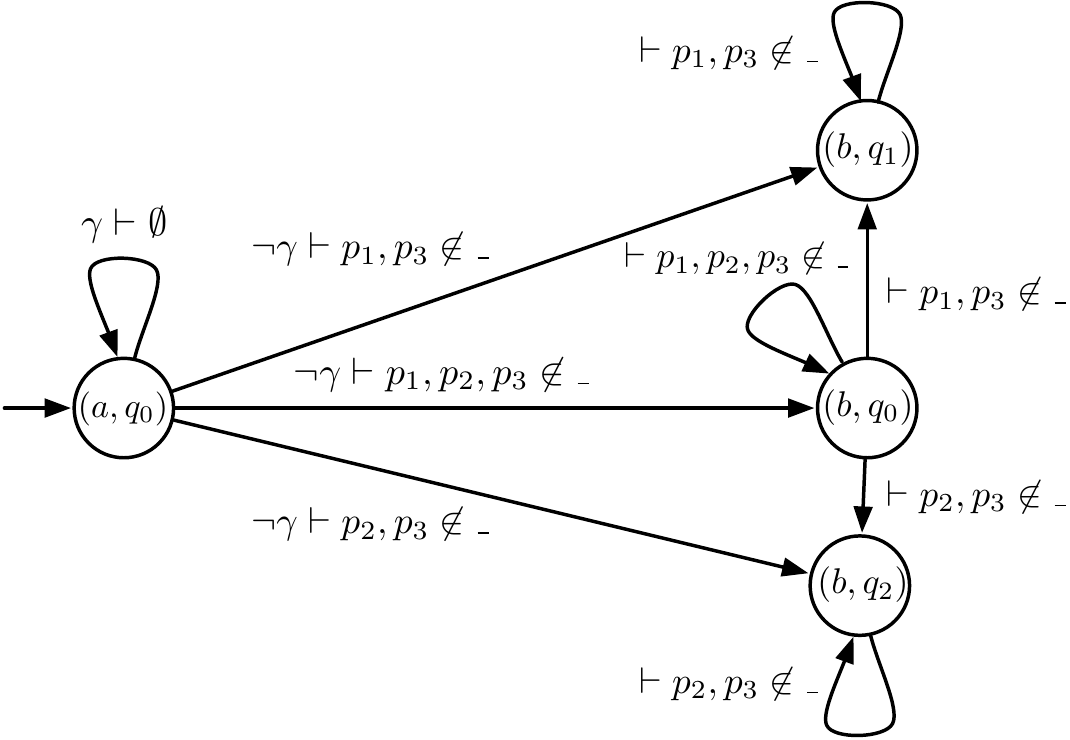}
	\caption{Composition $\mathcal{T}\parallel\mathrm{Excl}([t]_{\simeq_Y})$ for capability set $Y= \{p_1, p_2\}$}
	\label{fig:parallel_example_result}
\end{figure}

First, we construct $\mathrm{Excl}([t]_{\simeq_Y})$ in Figure~\ref{fig:construction_capset}, which characterises all tests that are not $Y$-set equivalent.
Then, we compose the strategies together, i.e.~$\mathcal{T}\parallel\mathrm{Excl}([t]_{\simeq_Y})$.
Figure~\ref{fig:parallel_example_result} depicts the result (with some simplifications), a strategy characterising all the same tests as $\mathcal{T}$ \emph{except} for those that are $Y$-set equivalent to $a$.
Note that $\emptyset\emptyset\emptyset PPQQQQ$ can no longer be derived as there are no paths allowing both $p_1$ and $p_2$; however, a test such as $\emptyset\emptyset PPPPP$ can still be derived, as this is not $Y$-set equivalent to $t$ (and thus is `different' under this equivalence class).
Note that the capability $p_3$ was forbidden in $\mathcal{T}$, and remains forbidden in the composition.

\begin{theorem}[Composition]\label{thm:composition}
	\emph{Let $\mathcal{P}$ denote a CPS, $\mathcal{T}$ a strategy, and $t$ a successful test on $\mathcal{P}$ such that $t \in L(\mathcal{T})$.
	For every equivalence class $\sim$ and successful test \mbox{$t'$ on $\mathcal{P}$:}}
	\[ \pi_{t'} \in L(\mathcal{T}\parallel\mathrm{Excl}([t]_{\sim})) \ \ \text{\emph{iff}}\ \ \pi_{t'} \in L(\mathcal{T}) \wedge t' \not\in [t]_\sim \]
	%\qed
\end{theorem}

%\noindent A proof is given in our appendix~\cite{Double-Blind-Appendix}.

\noindent A proof is given in the Appendix.

\section{Finding Different Tests by Guided Fuzzing}\label{sec:implementation}

Next, we show how our formalism can be combined with guided fuzzing approaches to automatically derive a set of mutually \emph{non-equivalent} tests from an initial strategy.

Our overall fuzzing approach is summarised in Figure~\ref{fig:overview_of_approach} and Algorithm~\ref{alg:overall_algorithm}: given some CPS $\mathcal{P}$ (e.g.~the SWaT testbed), a strategy $\mathcal{T}$, a goal $\gamma_g$ (e.g.~cause an overflow in tank $\mathtt{T101}$), and an equivalence class $\sim$ (e.g.~strong capability-set equivalence), our fuzzing algorithm will find and return a set of tests that are all mutually non-equivalent under $\sim$.

\begin{algorithm}[!t]
\caption{Overall Guided Fuzzing Approach}\label{alg:overall_algorithm}
\scriptsize
\KwIn{CPS $\mathcal{P}$; strategy $\mathcal{T}$; goal $\gamma_g$, equivalence class~$\sim$}
\KwOut{Set of distinct $tests$ for $\gamma_g$}
Let $tests := \{\}$; \\
\Repeat{\emph{timeout}}
{
	
	Derive a sequence of transitions $w = tr_1tr_2\cdots tr_n$ from $\mathcal{T}$ where $tr_i = (r_{i-1}, (\gamma_i, \_ == Y_i), r_i)$; [Alg.~\ref{alg:derive_attack}]\label{alg:line:derive_plan}\\
	
	Let $q_0, x_0$ denote the current control/physical states of $\mathcal{P}$;\\
	Let $t := (q_0, x_0, r_0)$;\\
	Let $i := 1$;\\	
	
	\While{$i \leq |w|$}
	{	
		Utilise capabilities $Y_i$ on $\mathcal{P}$ for $\tau_\mathcal{P}$ seconds;\\
		Let $q_i, x_i$ denote the current control/physical states of $\mathcal{P}$;\\
		\If{$\gamma_i^{x_i}$}
		{
			$t := t \rightarrow_{Y_i} (q_i, x_i, r_i)$;\\
			$i := i+1$;\\
		}
		\Else
		{
			$i := |w| + 1$;\\
		}
	}

	\If{$t$ \emph{is successful for} $\gamma_g$}
	{
		\If{\emph{causal fuzzing enabled}}
		{
			$t := t_\mathrm{min}$; [Alg.~\ref{alg:causal_events}]\label{alg:line:enable_causal}\\
		}
		$tests := tests \cup t$; \\
		$\mathcal{T} := \mathcal{T} \parallel \mathrm{Excl}([t]_\sim)$;\label{alg:line:update_strategy}\\
	}
}
\Return $tests$;
\end{algorithm}

The first step of the algorithm is to \emph{plan} a finite sequence of transitions (a `walk') from strategy $\mathcal{T}$ to fire on CPS $\mathcal{P}$.
This plan can be derived in a number of ways (e.g.~randomly), but we use a more intelligent approach which we describe separately in Section~\ref{sec:searching_for_an_attack}.
We aim to derive a sequence of transitions from $\mathcal{T}$ \emph{predicted} to achieve the goal $\gamma_g$ on $\mathcal{P}$.

Once a plan is obtained, the fuzzer proceeds to \emph{fire} the transitions one-by-one, utilising the capabilities of each transition for the given time interval $\tau_\mathcal{P}$.
The test $t$ consists of the sequence of control and physical states of $\mathcal{P}$ that result from firing those transitions, with the sequence terminating when either the final transition of the plan is fired, or when the next transition has a sensor condition that the physical state does not satisfy.
The test $t$ is \emph{successful} if $\gamma_g$ is true of the final physical state, in which case $t$ is added to the set of $tests$ returned by the fuzzing algorithm.
(If causal fuzzing is enabled, non-causal capabilities are pruned first---see Section~\ref{sec:approximation}.)

Before the main loop of the algorithm is repeated, the test strategy $\mathcal{T}$ is updated according to the algorithms of Section~\ref{sec:ident_classes} so as to ensure that the next test identified is not equivalent.

\subsection{Searching for a Potential Test}
\label{sec:searching_for_an_attack}

A key step of our overall fuzzing algorithm (Alg.~\ref{alg:overall_algorithm}, Line~\ref{alg:line:derive_plan}) is deriving a walk from a strategy, i.e.~planning a finite sequence of transitions from a strategy $\mathcal{T}$ to fire on the real system.
In general, strategies are nondeterministic, so there are many potential sequences: some states may have multiple outgoing transitions, and there may be $\mt{\_}$-labelled transitions in which different combinations of capabilities can be used.
As the search space is large, we propose an intelligent approach that searches for a walk \emph{predicted} to achieve the test goal $\gamma_g$.

Our approach for generating a plan is given in Algorithm~\ref{alg:derive_attack}.
Given a CPS $\mathcal{P}$, strategy $\mathcal{T}$, objective function $f_{\gamma_g}$, and prediction model $M_\mathcal{P}$, the algorithm returns a sequence of transitions (a `walk') in $\mathcal{T}$ that is predicted to maximise $f_{\gamma_g}$.
Here, $f_{\gamma_g}\!: \mathbb{S}_\mathcal{P} \rightarrow \mathbb{R}$ is some function that maps the sensor readings of $\mathcal{P}$ to a number, such that the larger that number is, the `closer' the physical state is to satisfying $\gamma_g$.
%While these objective functions must be defined, for our purposes, they are typically quite simple (see e.g.~\cite{Double-Blind-Appendix}).
While these objective functions must be defined, for our purposes, they are typically quite simple (see the Appendix).
Furthermore, the algorithm requires a model $M_\mathcal{P}\!:\mathbb{S}_\mathcal{P} \times \mathbb{A}_\mathcal{P}\rightarrow \mathbb{S}_\mathcal{P}$ that can predict how the current sensor readings will evolve over a given time period $\tau_\mathcal{P}$ relative to some intended actuator commands.
This model could be in the form of a simulator for $\mathcal{P}$, or it could be a machine learning model (e.g.~a neural network) that was trained on some data logs~\cite{Goh_et-al17a,Chen-Poskitt-et_al19a,Chen-Xuan-Poskitt-et_al20a}.

\begin{algorithm}[!t]
\caption{Find a Potential Test from a Strategy}\label{alg:derive_attack}
\scriptsize
\KwIn{CPS $\mathcal{P}$; strategy $\mathcal{T}$; objective function $f_{\gamma_g}$; prediction model $M_\mathcal{P}$}
\KwOut{Sequence of transitions $w$}

Let $walks := \langle \rangle$; [empty sequence]\\
Let $scores := \langle \rangle$;\\

\Repeat{\emph{timeout}}
{
	From $i_\mathcal{T}$, generate finite random walk $tr_1tr_2\cdots tr_n$ over transitions $tr_i$ in $\mathcal{T}$; \\
	Replace $\varphi_i$ in each $tr_i$ with $Y_i$ for some $Y_i$ satisfying $\varphi_i$;\\
	$walks := walks.\mathrm{append}(tr_1tr_2\cdots tr_n)$;\\
}

Let $\overline{S}_0, \overline{A}_0$ denote $\mathcal{P}$'s current sensor/actuator values;\\

\ForEach{$w \in walks$}
{
	%Let $\overline{S}, \overline{A} := \overline{S}_0, \overline{A}_0$;\\
    Let $\overline{S} := \overline{S}_0$;\\
	Let $i = 1$;\\
	\While{$i < |w|$}
	{
        Let $\overline{A}_i$ denote the actuator commands resulting from $Y_i$;\\
		%$\overline{A} := \overline{A}^{Y_i}$;\\
		$\overline{S} := M_\mathcal{P}(\overline{S}, \overline{A}_i)$;\\
		$i := i + 1$;\\
	}
	$scores.\mathrm{append}(f_{\gamma_g}( M_\mathcal{P}(\overline{S}, \overline{A}_{|w|}) )) $; \\

}
Select a walk $w$ from $walks$ using \emph{roulette wheel selection} with corresponding scores in $scores$;\\
\Return walk $w$;
\end{algorithm}

Intuitively, the algorithm begins by generating a large number of completely random finite walks through the strategy $\mathcal{T}$. 
Following this, the algorithm processes each random walk in turn, \emph{predicting} the sensor readings that would result from actually firing the transitions relative to the current state of the CPS.
This final prediction is achieved by applying the model $M_\mathcal{P}$ to each transition in turn, and then applying the objective function $f_{\gamma_g}$ to the final state.
The higher the number, the `closer' the predicted effects of the walk are to achieving the testing goal.
Rather than simply returning the walk with the highest score, we choose one using roulette wheel selection~\cite{Goldberg89a} in order to ensure diversity in the tests.% explored.

\subsection{Causally Different Tests}
\label{sec:approximation}

Thus far, our equivalence classes have not taken into account the \emph{causality} of capability usages.
For example, a test that uses capabilities $Y$ and a test that uses capabilities $Y \cup \{p\}$ (with $p\not\in Y$) would be considered different under strong capability-set equivalence, even if $p$ is totally irrelevant to the test's success (e.g.~it modifies an actuator in an unrelated stage).

Our solution here is for equivalence classes to consider \emph{only} those events that were causal for the test's success.
As computing the exact causal set is NP-complete in general~\cite{Beer-et_al09a,Beer-et_al12a}, we propose an algorithm to efficiently approximate it.
In particular, given a successful test, our algorithm prunes capability usages from it that are non-causal.
To assess causality, we analyse counterfactual conditionals of the form ``if $p$ did not occur, then $\gamma_g$ would not be achieved'', by systematically replaying the test with individual capabilities removed.
Depending on the CPS $\mathcal{P}$ involved, this replay can be performed on the system itself, or by using the prediction model $M_\mathcal{P}$ from the previous steps (i.e.~either a learnt model or a high-fidelity simulator).

\begin{figure}[!t]
	\centering
	\includegraphics[width=0.6\linewidth]{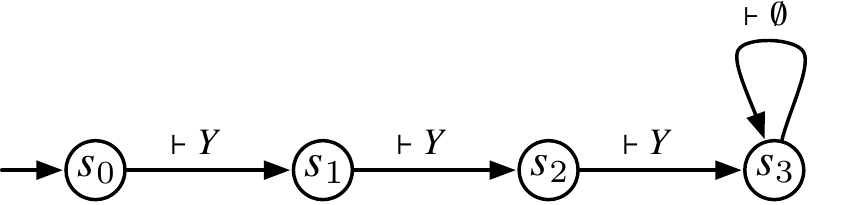}
	\caption{Example of a replay strategy}
	\label{fig:replay_strategy}
\end{figure}

We can formalise this notion of `replaying' a test as a strategy. Given a test $t$ with capability history $\pi_t = Y_1Y_2\cdots Y_n$, the \emph{replay strategy} for $\pi_t$ is $\mathcal{T}_{\pi_t} = (\{s_0, \dots, s_n\}, \{tr_1, \dots, tr_n, tr_\emptyset\}, s_0)$, where each $tr_i = (s_{i-1}, (\mathrm{true}, \_==Y_i), s_i)$ and $tr_\emptyset = (s_n, (\mathrm{true}, \_==\emptyset), s_n)$.
Figure~\ref{fig:replay_strategy}, for example, depicts the replay strategy constructed from a history $\pi_t = YYY$ for $Y = \{ [ \mathtt{P_{101}}, \mathtt{on}], [ \mathtt{MV_{101}}, \mathtt{close}] \}$.

The following definition formalises our notion of causality in a successful test for $\gamma_g$.
In particular, the usage of a capability $p$ across a given range of positions is \emph{causal} if removing $p$ from those positions would mean the resulting test no longer satisfies $\gamma_g$.
This is assessed using replay strategies in which $p$ has been removed.

\begin{definition}[Causal capability]\label{def:causal_capability}\rm
	Let $t = (q_0, x_0, r_0) \rightarrow_{Y_1} \cdots$ denote a successful test for $\gamma_g$ with capability history $\pi_t = Y_1Y_2\cdots Y_n$. Given a capability $y$ and indices $k,l$ such that $1 \leq k \leq l \leq n$, we say that $y$ is \emph{causal in} $\pi_t$ \emph{from} $k$ \emph{to} $l$ if there does not exist a successful test for $\gamma_g$ from $q_0,x_0$ using replay strategy $\mathcal{T}_{\pi_t\setminus\langle y, k, l \rangle}$.
	If such a test from $q_0,x_0$ exists, we call it the \emph{counterexample}.
	Note that the sequence $\pi_t\setminus\langle y, k, l \rangle$ is obtained from $\pi_t$ by substituting $Y_i\setminus\{y\}$ for $Y_i$ across $k \leq i \leq l$.    
	\qed
\end{definition}

With these concepts defined, Algorithm~\ref{alg:causal_events} describes a linear-time approximation procedure that takes a successful test $t$ for $\gamma_g$ and returns a successful test $t_\mathrm{min}$ for $\gamma_g$ in which non-causal capability usages have been removed.
The algorithm operates by repeatedly identifying a maximal subsequence of equal, repeated capability set usages, and replaying the test with a specific capability removed from each of the sets in that sub-sequence. For example, for a sequence $PPPQ$ with $p\in P$, the algorithm may replay the sequence $P'P'P'Q$ where $P'=P\setminus \{p\}$.
If the test goal is still met, the capability usage was not causal and can be pruned; if it is not, then it is recorded as causal and other subsequences of capabilities are tested.
When no remaining capability can be pruned, the test is returned.

\begin{algorithm}[!t]
\caption{Approximating the Causal Capabilities}\label{alg:causal_events}
\scriptsize
\KwIn{Successful test $t$ for $\gamma_g$} 
\KwOut{Pruned successful test $t_\mathrm{min}$ for $\gamma_g$}

Let $t_\mathrm{min} := t$; \\
Let $causal := \{\}$; \\

\Repeat{\emph{Line~\ref{alg:line:pick} fails}}      
{
	Pick $y,k,l$ where $\pi_{t_\mathrm{min}}[k..l]$ is a maximal slice of equal capability sets, $y$ is a capability in the set, and $\langle y, k, l \rangle \notin causal$; \label{alg:line:pick} \\

	\If{$y$ \emph{is causal in} $\pi_{t_\mathrm{min}}$ \emph{from} $k$ \emph{to} $l$}
	{
		$causal := causal \cup \{\langle y, k, l \rangle\}$; \\

	}
	\Else
	{

		Let $t_\mathrm{min}$ denote the \emph{counterexample}; [Def.~\ref{def:causal_capability}]\\

	}
}
\Return $t_\mathrm{min}$;

\end{algorithm}

\begin{proposition}[Over-approximation of causal capabilities]
	\emph{Given a successful test $t$ for $\gamma_g$, Algorithm~\ref{alg:causal_events} returns a test containing an over-approximation of its causal capabilities.}
\qed
\end{proposition}

\noindent This is evident given that capabilities are only removed if there exist counterexamples to them being causal.
The algorithm's complexity is linear in the number of events in the trace.

Finally, to ensure that Algorithm~\ref{alg:overall_algorithm} finds \emph{causally different} tests, we replace $t$ with $t_\mathrm{min}$ (Line~\ref{alg:line:enable_causal}), and then use $\mathrm{CSet}(\pi_{t_\mathrm{min}})$-equivalence as our equivalence class when updating $\mathcal{T}$ (Line~\ref{alg:line:update_strategy}).
This ensures that any other tests identified by the algorithm are not based on the same causal manipulations.

\section{Evaluation on SWaT} \label{sec:evaluation}

In this section, we evaluate the effectiveness of our formalisation and guided fuzzing approach with respect to SWaT, a real-world testbed representative of ICSs in the water domain. In particular, we ask:

\begin{description}
	\item[\textbf{RQ1:}] How many \emph{different tests} can our approach automatically find for each testing goal? 
	\item[\textbf{RQ2:}] Does causal fuzzing effectively prune minor permutations of otherwise equivalent tests? 
	\item[\textbf{RQ3:}] How long does it take to find these tests? 
	\item[\textbf{RQ4:}] Does causal fuzzing generate a more diverse set of tests than the SWaT benchmark? 
\end{description}

\noindent Through this evaluation, we aim to judge whether our formal approach can really scale to a complex and real-world critical infrastructure system from the water domain.

%Our programs are available online~\cite{Supplementary-Material}, and details of the implementation of our algorithms for SWaT are available in the Appendix.

Our programs are available online~\cite{Supplementary-Material}, and details of the implementation of our algorithms for SWaT are available in the Appendix.

\textbf{\emph{Experiment \#1: RQ1.}} Our first experiment assessed how many \emph{different} tests for SWaT our algorithms can find according to two of our equivalence classes: strong capability-set equivalence, and strong capability-order equivalence.
These classes only take into account patterns of capability usages: they do not consider the causality of individual manipulations, the importance of which is explored in our second experiment.

To achieve this, we initialised our implementation of the algorithm on a `universal' test strategy, i.e.~a single node incident to a transition permitting any capability usages.
We ran our program for one minute per goal (i.e.~per unsafe state) and equivalence class, using the SWaT simulator as our prediction model for Algorithm~\ref{alg:derive_attack}, randomly generating an initial configuration each time within normal operational ranges.
For test goals concerning all sensors except the LITs, we set the time interval $\tau_\mathcal{P}$ as 15s, as the changes take effect very quickly (no longer than 15s in preliminary experimentation).
Given the size of tanks, it takes somewhat longer to effect changes on the LIT readings.
Thus, we chose a separate time interval of 10 minutes for test goals concerning the LITs.
Note that we do not mix different time intervals in a single test: the value of $\tau_\mathcal{P}$ is selected according to the test goal, and applied for each capability usage in the test. (Future work could generalise this in the formal model, perhaps facilitating equivalence classes defined with respect to time.)

This value of $\tau_\mathcal{P}$ was determined in a simple pre-study: we randomly generated an initial configuration within the SWaT simulator's normal operational ranges, searched for a combination of actuators predicted to drive the LIT sensor to an unsafe state, then set that combination in the simulator and ran it 100 times, recording the median time taken to reach the unsafe state.
We found that no sensor required more than 30 minutes, and with three transitions per walk (see the Appendix), this suggested a time interval of 10 minutes.
Note that as part of this pre-study, we recreated one run per test goal on the physical testbed to validate the time intervals identified by experimenting on the simulator.

Our results are given in the first two rows of Table~\ref{tab:results:part1}.
We found that our implementation was able to quickly generate large numbers of effective tests (as predicted by the simulator) for every test goal, with each test distinct according to the equivalence class definition used.
As mentioned before, these equivalence classes are \emph{non-causal}, defined only over patterns of the capabilities present, meaning that minor permutations of the same test are not filtered out.
Our next experiment considers whether our causal-aware fuzzing solves this problem.

\begin{table*}[!t]

    \caption{Results: number of different tests (according to equivalence class) found for each testing goal}
    \label{tab:results:part1}
	%\centering
	%\includegraphics[width=\linewidth]{figs/temp_results_table.pdf}
 \resizebox{\linewidth}{!}{
 \begin{tabular}{|c|c||ccc|ccc|ccc|cccccc|c|}
	\cline{3-18}
	%\multirow{2}{*}{Fuzzing Configuration} 
	\multicolumn{2}{c|}{} & \multicolumn{3}{c|}{Tanks (High)} & \multicolumn{3}{c|}{Tanks (Low)} &\multicolumn{3}{c|}{Flow (High)} &\multicolumn{6}{c|}{Flow (Low)} & Pr.~(L) \\ %&\multicolumn{2}{c||}{Pressure} \\ % \multicolumn{1}{c}{}
	\cline{3-18}
	\multicolumn{2}{c|}{} &  \rotatebox{90}{LIT101} & \rotatebox{90}{LIT301} & \rotatebox{90}{LIT401} & \rotatebox{90}{LIT101} & \rotatebox{90}{LIT301} & \rotatebox{90}{LIT401} &  \rotatebox{90}{FIT101} & \rotatebox{90}{FIT201} & \rotatebox{90}{FIT601} &  \rotatebox{90}{FIT101} & \rotatebox{90}{FIT201} & \rotatebox{90}{FIT301} & \rotatebox{90}{FIT401} & \rotatebox{90}{FIT501} & \rotatebox{90}{FIT601} & \rotatebox{90}{DPIT301} \\ % & DPIT301 (High) & PIT501 (Low) & XXX & XXX & XXX \\
	%\hline\hline
	\hhline{--================}
	\emph{Non-} & Strong capability-set equivalence & 34,019 & 30,034 & 38,890 & 23,659 & 29,853 & 37,013 & 42,355 & 51,902 & 7,713 & 41,091 & 54,109 & 41,023 & 41,001 & 16,265 & 41,045 & 56,834$^\ast$     \\
	%\cline{2-10}
	\emph{Causal} & Strong capability-order equivalence & 40,199 & 28,100 & 34,191 & 19,908 & 27,118 & 44,213 &54,239 & 56,122 & 10,902 & 46,124 & 54,213 & 49,008 & 42,123 & 23,455 & 44,214 & 59,901$^\ast$   \\
	\hline
	%\cline{2-10}
	\emph{Causal} & Causal-set equivalence & \textbf{3} & \textbf{6} & \textbf{2} & \textbf{2} & \textbf{2} & \textbf{4} & 1 & 1 & 1 & 1 & \textbf{2} & \textbf{2} & 1 & 1 & 1 & \textbf{3}        \\
	%\cline{2-10}
	\hline%\hline
	\emph{Other} & SWaT dataset/benchmark & 2 & 1 & --- & 1 & --- & 1 & --- & --- & --- & --- & --- & --- & --- & --- & --- & ---  \\
	\hline
\end{tabular}}

\end{table*}

\textbf{\emph{Experiment \#2: RQ2--3.}} Our second experiment assesses how many \emph{causally different} tests for SWaT our algorithm can find, and how much additional time our causality approximation adds in comparison to Experiment \#1.
In particular, we searched for new tests on the basis of our $Y$-set equivalence class, where $Y$ is taken to be the \emph{set of causal events} in the test (as approximated by Algorithm~\ref{alg:causal_events}).
As for our previous experiment, we started with a `universal' test strategy that could derive any possible combination of capabilities.

First, we used our program to search for a single test $t$ predicted to achieve the test goal.
We then applied Algorithm~\ref{alg:causal_events} to approximate the causal capabilities, pruning out all others to get $t_\mathrm{min}$.
For all non-LIT goals, we performed this approximation directly on the testbed.
Due to the time requirements for LIT goals, we implemented those causality tests using the simulator (on a random initial state within normal operational ranges), \emph{except} for the final minimised sequence which was tested and verified on the real system.
Next, we parallel composed the strategy with $\mathrm{Excl}([t]_{\simeq_Y})$, where $Y$ is taken as $\mathrm{CSet}(\pi_{t_\mathrm{min}})$.
We repeated this process for the next test found, which was guaranteed by our formalisation not to contain this causal set of capabilities.

Our results are presented in the third row of Table~\ref{tab:results:part1}, and reflect the total number of tests that could be found based on different causality sets of capabilities.
For each non-LIT goal, the causality approximation algorithm took up to an hour per test to complete on the testbed.
For each LIT goal, in which the final step was run on the testbed but others on the simulator, the approximation took 10 minutes per test.

The number of different tests represents a significant reduction from those of the non-causal experiment, and illustrates the power of causality-based equivalence classes in reducing the search space.
In fact, we found that $100\%$ of the tests from the previous experiment utilised at least one of the causality sets identified in this one, meaning that they were minor permutations of the key manipulations that made the test succeed.
The only exception was DPIT301-Low ($\ast$), for which $97\%$ of the tests utilised one of the three causality sets.
We believe that this was due to a poorer prediction model for the sensor in our simulator, although the $3\%$ of tests still pushed the sensor reading towards the testing goal.

%The approximated causality sets are given online~\cite{Double-Blind-Appendix}.
The approximated causality sets are given in the Appendix.
Upon manual inspection, we found these approximations to be exact in all cases.
We note that for some test goals, some causality sets are subsets of others (e.g.~for LIT301-High).
The reason for this is due to the initial configuration of the system that the test is launched in: some tests based on those causal sets will only succeed when the water tanks are at certain levels (e.g.~close to unsafe ranges).

\begin{center}
\noindent\fbox{%
    \parbox{0.9\linewidth}{%
        \small\emph{Our approach avoids generating causally equivalent tests for SWaT, pruning the test suite by 4 orders of magnitude, and is practically implemented with a linear-time approximation.}

    }
}
\end{center}

\noindent Note that for other systems, one may wish to generate multiple tests for each equivalence class. This could be achieved by triggering the strategy update only after certain thresholds (e.g.~after generating $n$ number of tests for a given class). Due to resource limitations, we do not evaluate this alternative approach for SWaT.

\textbf{\emph{Experiment \#3: RQ4.}} Finally, we assessed whether causal fuzzing led to tests that achieve better coverage/diversity than existing CPS fuzzers~\cite{Chen-Poskitt-et_al19a} and the SWaT benchmark~\cite{Goh-et_al16a,CPS-Datasets} (Section~\ref{sec:background}) by counting the number of tests that had the goal of driving a sensor into an unsafe range. 
We chose the latter as our benchmark as it is the most extensive publicly available dataset for the SWaT testbed, and has become established as a baseline in the critical infrastructure community (cited over 300 times since 2016, and downloaded by researchers in 82 countries~
\cite{CPS-Datasets}.)

As shown in Table~\ref{tab:results:part1}, none of the tests in the benchmark target the unsafe states of the FITs or DPIT, covering only the unsafe states of LITs. 
Even limited to LITs, our causal fuzzer still has more coverage/diversity: the benchmark does not contain tests for the LIT401~(High) and LIT301~(Low) ranges, and contains only 1--2 tests for the others.
We note, however, that some of the tests in the benchmark are incomparable to ours as they do not target the manipulation of a physical state, but rather just spoof a sensor (without physical effects); causal fuzzing focuses only on tests that cause physical changes.

With the CPS fuzzer of~\cite{Chen-Poskitt-et_al19a}, we are only able to find one causally distinct test per each of our 16 test goals.
This is explained by its search approach, which is always guided towards the `simplest' causal capabilities.
Thus, across these goals, causal fuzzing finds 106\% more causally different tests.

\begin{center}
\noindent\fbox{%
    \parbox{0.7\linewidth}{%
        \small\emph{Causal fuzzing found 106\% more causally different tests for SWaT than the most comparable CPS fuzzer.}
    }
}
\end{center}

\textbf{\emph{Threats to Validity.}}
While SWaT is a fully operational testbed, it is not as large as the plants it is based on, meaning our results may not scale-up (this is difficult to assess due to strict confidentiality).
Furthermore, due to practical constraints, our causality analysis (Algorithm~\ref{alg:causal_events}) is partially evaluated on the simulator (although the final results are all validated on the actual testbed).
As a result, the success of a test may depend on the starting state/time: occasionally, the system may not reach the unsafe state even for causal events, thus causal events might be removed.
This could be addressed with a weaker requirement for being causal, e.g.~an event could be considered causal as long as there is a physical state change towards the test goal.

We remark that our evaluation has focused on a real-world ICS in the water domain. While our formalisms and algorithms were designed with generality in mind, the results of these experiments may not generalise to different kinds of CPSs (e.g.~drones). This should be explored in future work.

\section{Related Work} \label{sec:related_work}

In this section, we highlight how our approach relates to other works involving fuzzing and test generation for CPSs and ICSs in particular. While fuzzing has been applied to these kinds of systems before, the aims are often quite different (e.g.~to find bugs in network protocols) in comparison to ours, which is to provide a formal causality-guided way of diversifying the test suite of an ICS.

Our approach was motivated by problems observed in the original `CPS fuzzers' developed in our previous work~\cite{Chen-Poskitt-et_al19a,Chen-Xuan-Poskitt-et_al20a}.
%The closest works to ours are the `CPS fuzzers' of Chen et al.~\cite{Chen-Poskitt-et_al19a,Chen-Xuan-Poskitt-et_al20a}.
For example, in~\cite{Chen-Poskitt-et_al19a}, a genetic algorithm is used to find actuator configurations of SWaT that were predicted to drive the system into unsafe physical states.
The search attempts to find \emph{some} test of this kind as quickly as possible, which in practice means that it will usually return the same test for a given goal (typically the `simplest' test, but perhaps with some random non-causal manipulations as a result of the optimisation).
Our new fuzzer avoids this problem by determining the causal manipulations and excluding the entire class of causally equivalent classes from the search space, ensuring that multiple causally different tests can be found for a given goal (see e.g.~Table~\ref{tab:results:part1}).
We also found more tests than the active learning-based fuzzer we proposed in~\cite{Chen-Xuan-Poskitt-et_al20a}, although a direct comparison here is less fair as it focused on manipulating low-level network packets (i.e.~bit-vectors) rather than high-level actuator commands.

Other CPS fuzzers are more domain-specific, focusing on finding bugs in code or protocols, rather than sensor/actuator manipulations.
CyFuzz~\cite{Chowdhury-Johnson-Csallner17a}, for example, tests CPS tool chain components that have been implemented in Simulink by randomly generating input models.
DeepFuzzSL~\cite{Shrestha-Chowdhury-Csallner20a} goes further by guiding the generation with a neural network.
Fuzzing has also been applied to specific protocols in CPSs, e.g.~network protocols in order to test their intrusion detection systems (e.g.~\cite{Vigna-et_al04a}).
Our work differs in that subsequent phases of test generation adapt to ensure causally equivalent classes of tests are excluded from the search.

Several works have applied the concept of causality for \emph{security analyses} of CPSs. In~\cite{DBLP:conf/fmics/MoradiASCST20}, Moradi et al.~employed a STRIDE model as a reference for classifying attacks.
While not explicitly built upon causality, their STRIDE model has a certain form of in-built causality. 
In~\cite{9141597}, Zhang et al.~applies causal models (defined based on maximum information coefficients and transfer entropy) to trace and detect ICS anomalies.
Outside the domain of CPSs, causality has been increasingly applied to analyse or explain complex systems including in AI~\cite{DBLP:conf/ecai/IbrahimKZKP20,DBLP:journals/cacm/Pearl19,DBLP:conf/icml/ForneyPB17}.
In contrast to these works, we use causality as a way to ensure diversity from our fuzzer.

Several authors have investigated the \emph{falsification of CPS specifications}. Given a CPS and a formal specification (e.g.~in some temporal logic), the idea is to search for counterexamples that violate the specification.
Tools such as S-TaLiRo~\cite{Annpureddy-et_al11a} use stochastic search methods to find simulation traces that minimise a global robustness metric.
Yamagata et al.~\cite{Yamagata-et_al20a} also attempt to minimise how robustly specifications are satisfied but drive the search using deep reinforcement learning.
The tests found by our work could be considered as similar to the counterexamples of these works, although otherwise our approach is quite different.
Our approach may combine to help find causally different ways of violating the same specification.

Outside of ICSs and CPSs, many fuzzers are available for testing software, e.g.~\cite{Zalewski,Cha-Woo-Brumley15a,Boehme-et_al17a,DBLP:conf/uss/Chen0MLWZJS19,DBLP:conf/icse/NguyenP0L020}.
A common approach to improve test generation is to specify the class of valid inputs as a context-free grammar~(CFG)~\cite{Godefroid-Kiezun-Levin08a,Zeller-et_al19a}.
While it would be feasible to specify our `strategies' as CFGs, and use CFG-guided fuzzers to generate tests, it may be difficult to exclude equivalence classes of tests from a CFG in general (especially as this requires handling language complements).
It is also worth noting that while CFGs typically express (single-step) software inputs, our strategies express (multi-step) sequences of system manipulations that are evaluated by observing their effects on execution traces.
Though our approach was motivated by a specific application in diversifying ICS test suites, it may general enough to adapt for fuzzing other types of software in which communication can be manipulated, e.g.~distributed software systems.

Johnson et al.~\cite{Johnson-Brun-Meliou20a} proposed `Causal Testing', an approach that helps Java developers to identify causal information associated with a failed test case by fuzzing with minimally different inputs that do not exhibit the faulty behaviour.
Our approach differs in that they use causality to \emph{explain} an existing failure, while we use it to diversify the test suite.
%whereas we adapt the existing strategy to generate different tests.

\section{Conclusion} \label{sec:conclusion}

Inspired by the limited diversity of existing test benchmarks for ICSs such as SWaT, we developed a formal guided fuzzing approach that allows us to systematically generate causally different tests with respect to an equivalence class and a given goal.
By focusing on traces of sensor/actuator manipulations, our approach can be applied without having to construct any mathematical models of the targeted system's programs or physics.
We assessed the utility of our approach by implementing it in a fuzzer for a real-world water treatment testbed, finding that it was able to identify multiple tests that were successful for \emph{causally different} reasons.

In future work, we would like to explore practical ways of reducing the overhead of our online strategy update algorithms. For example, as an alternative to excluding classes of tests by construction, a guided fuzzer may be able to incorporate test diversity (i.e.~being outside known equivalence classes) as a part of the `reward' function when searching for new tests.
We will also determine whether our approach can be applied to fuzzing different kinds of CPSs (e.g.~drones), or other software/systems in which the execution trace can be influenced by intercepting some communication channel.
Finally, we will explore its applicability to model checking (e.g.~searching for different counterexamples) and compositional verification (i.e.~verifying the system under the presence of different test strategies).

\section*{Acknowledgements}
    We are grateful to our anonymous referees who have helped immensely in improving the presentation and positioning of this paper.
    We would also like to thank Alexander Pretschner and Eric Rothstein-Morris for insightful discussions at the outset of this work.
    This research / project is supported by the National Research Foundation, Singapore, under its National Satellite of Excellence Programme ``Design Science and Technology for Secure Critical Infrastructure'' (Award Number: NSoE\_DeST-SCI2019-0008).
    Any opinions, findings and conclusions or recommendations expressed in this material are those of the author(s) and do not reflect the views of National Research Foundation, Singapore.

\bibliographystyle{IEEEtran}

\bibliography{references}

% Generated by IEEEtran.bst, version: 1.14 (2015/08/26)
\begin{thebibliography}{10}
\providecommand{\url}[1]{#1}
\csname url@samestyle\endcsname
\providecommand{\newblock}{\relax}
\providecommand{\bibinfo}[2]{#2}
\providecommand{\BIBentrySTDinterwordspacing}{\spaceskip=0pt\relax}
\providecommand{\BIBentryALTinterwordstretchfactor}{4}
\providecommand{\BIBentryALTinterwordspacing}{\spaceskip=\fontdimen2\font plus
\BIBentryALTinterwordstretchfactor\fontdimen3\font minus
  \fontdimen4\font\relax}
\providecommand{\BIBforeignlanguage}[2]{{%
\expandafter\ifx\csname l@#1\endcsname\relax
\typeout{** WARNING: IEEEtran.bst: No hyphenation pattern has been}%
\typeout{** loaded for the language `#1'. Using the pattern for}%
\typeout{** the default language instead.}%
\else
\language=\csname l@#1\endcsname
\fi
#2}}
\providecommand{\BIBdecl}{\relax}
\BIBdecl

\bibitem{Inoue-et_al17a}
J.~Inoue, Y.~Yamagata, Y.~Chen, C.~M. Poskitt, and J.~Sun, ``Anomaly detection
  for a water treatment system using unsupervised machine learning,'' in
  \emph{Proc.\ {IEEE} International Conference on Data Mining Workshops (ICDMW
  2017)}.\hskip 1em plus 0.5em minus 0.4em\relax IEEE, 2017, pp. 1058--1065.

\bibitem{Aoudi-et_al18a}
W.~Aoudi, M.~Iturbe, and M.~Almgren, ``Truth will out: Departure-based
  process-level detection of stealthy attacks on control systems,'' in
  \emph{Proc.\ {ACM} {SIGSAC} Conference on Computer and Communications
  Security (CCS 2018)}.\hskip 1em plus 0.5em minus 0.4em\relax {ACM}, 2018, pp.
  817--831.

\bibitem{Kravchik-Shabtai18a}
M.~Kravchik and A.~Shabtai, ``Detecting cyber attacks in industrial control
  systems using convolutional neural networks,'' in \emph{Proc.\ Workshop on
  Cyber-Physical Systems Security and PrivaCy (CPS-SPC 2018)}.\hskip 1em plus
  0.5em minus 0.4em\relax {ACM}, 2018, pp. 72--83.

\bibitem{Lin-et_al18a}
Q.~Lin, S.~Adepu, S.~Verwer, and A.~Mathur, ``{TABOR:} {A} graphical
  model-based approach for anomaly detection in industrial control systems,''
  in \emph{Proc.\ Asia Conference on Computer and Communications Security
  (AsiaCCS 2018)}.\hskip 1em plus 0.5em minus 0.4em\relax {ACM}, 2018, pp.
  525--536.

\bibitem{Carrasco-Wu19a}
M.~A.~M. Carrasco and C.~Wu, ``An unsupervised framework for anomaly detection
  in a water treatment system,'' in \emph{Proc.\ {IEEE} International
  Conference On Machine Learning And Applications ({ICMLA} 2019)}.\hskip 1em
  plus 0.5em minus 0.4em\relax {IEEE}, 2019, pp. 1298--1305.

\bibitem{Adepu-et_al20a}
S.~Adepu, F.~Brasser, L.~Garcia, M.~Rodler, L.~Davi, A.~Sadeghi, and S.~A.
  Zonouz, ``Control behavior integrity for distributed cyber-physical
  systems,'' in \emph{Proc.\ {ACM/IEEE} International Conference on
  Cyber-Physical Systems ({ICCPS} 2020)}.\hskip 1em plus 0.5em minus
  0.4em\relax {IEEE}, 2020, pp. 30--40.

\bibitem{Ahmed-et_al20a}
C.~M. Ahmed, A.~P. Mathur, and M.~Ochoa, ``{NoiSense Print}: Detecting data
  integrity attacks on sensor measurements using hardware-based fingerprints,''
  \emph{ACM Transactions on Privacy and Security}, vol.~24, no.~1, 2020.

\bibitem{Gu-et_al18a}
Q.~Gu, D.~Formby, S.~Ji, H.~Cam, and R.~A. Beyah, ``Fingerprinting for
  cyber-physical system security: Device physics matters too,'' \emph{{IEEE}
  Security {\&} Privacy}, vol.~16, no.~5, pp. 49--59, 2018.

\bibitem{Kneib-Huth18a}
M.~Kneib and C.~Huth, ``Scission: Signal characteristic-based sender
  identification and intrusion detection in automotive networks,'' in
  \emph{Proc.\ {ACM} {SIGSAC} Conference on Computer and Communications
  Security (CCS 2018)}.\hskip 1em plus 0.5em minus 0.4em\relax {ACM}, 2018, pp.
  787--800.

\bibitem{Adepu-Mathur18b}
S.~Adepu and A.~Mathur, ``Distributed attack detection in a water treatment
  plant: Method and case study,'' \emph{IEEE Transactions on Dependable and
  Secure Computing}, vol.~18, no.~1, pp. 86--99, 2021.

\bibitem{Chen-Poskitt-Sun18a}
Y.~Chen, C.~M. Poskitt, and J.~Sun, ``Learning from mutants: Using code
  mutation to learn and monitor invariants of a cyber-physical system,'' in
  \emph{Proc.\ {IEEE} Symposium on Security and Privacy (S{\&}P 2018)}.\hskip
  1em plus 0.5em minus 0.4em\relax {IEEE} Computer Society, 2018, pp. 648--660.

\bibitem{Choi-et_al18a}
H.~Choi, W.~Lee, Y.~Aafer, F.~Fei, Z.~Tu, X.~Zhang, D.~Xu, and X.~Xinyan,
  ``Detecting attacks against robotic vehicles: {A} control invariant
  approach,'' in \emph{Proc.\ {ACM} {SIGSAC} Conference on Computer and
  Communications Security ({CCS} 2018)}.\hskip 1em plus 0.5em minus 0.4em\relax
  {ACM}, 2018, pp. 801--816.

\bibitem{Giraldo-et_al18a}
J.~Giraldo, D.~I. Urbina, A.~Cardenas, J.~Valente, M.~A. Faisal, J.~Ruths,
  N.~O. Tippenhauer, H.~Sandberg, and R.~Candell, ``A survey of physics-based
  attack detection in cyber-physical systems,'' \emph{{ACM} Computing Surveys},
  vol.~51, no.~4, pp. 76:1--76:36, 2018.

\bibitem{Yoong-et_al21a}
C.~H. Yoong, V.~R. Palleti, R.~R. Maiti, A.~Silva, and C.~M. Poskitt,
  ``Deriving invariant checkers for critical infrastructure using axiomatic
  design principles,'' \emph{Cybersecurity}, vol.~4, no.~1, p.~6, 2021.

\bibitem{Chen-Poskitt-et_al19a}
Y.~Chen, C.~M. Poskitt, J.~Sun, S.~Adepu, and F.~Zhang, ``Learning-guided
  network fuzzing for testing cyber-physical system defences,'' in \emph{Proc.\
  IEEE/ACM International Conference on Automated Software Engineering (ASE
  2019)}.\hskip 1em plus 0.5em minus 0.4em\relax {IEEE} Computer Society, 2019,
  pp. 962--973.

\bibitem{Chen-Xuan-Poskitt-et_al20a}
Y.~Chen, B.~Xuan, C.~M. Poskitt, J.~Sun, and F.~Zhang, ``Active fuzzing for
  testing and securing cyber-physical systems,'' in \emph{Proc.\ ACM SIGSOFT
  International Symposium on Software Testing and Analysis (ISSTA 2020)}.\hskip
  1em plus 0.5em minus 0.4em\relax ACM, 2020, pp. 14--26.

\bibitem{Wijaya-Aniche-Mathur20a}
H.~Wijaya, M.~Aniche, and A.~Mathur, ``Domain-based fuzzing for supervised
  learning of anomaly detection in cyber-physical systems,'' in \emph{Proc.\
  International Workshop on Engineering and Cybersecurity of Critical Systems
  (EnCyCriS 2020)}.\hskip 1em plus 0.5em minus 0.4em\relax {ACM}, 2020, pp.
  237--244.

\bibitem{Kim-et_al21a}
H.~Kim, M.~O. Ozmen, A.~Bianchi, Z.~B. Celik, and D.~Xu, ``{PGFUZZ}:
  Policy-guided fuzzing for robotic vehicles,'' in \emph{Proc.\ Annual Network
  and Distributed System Security Symposium ({NDSS} 2021)}.\hskip 1em plus
  0.5em minus 0.4em\relax The Internet Society, 2021.

\bibitem{Goh-et_al16a}
J.~Goh, S.~Adepu, K.~N. Junejo, and A.~Mathur, ``A dataset to support research
  in the design of secure water treatment systems,'' in \emph{Proc.\
  International Conference on Critical Information Infrastructures Security
  (CRITIS 2016)}, ser. LNCS, vol. 10242.\hskip 1em plus 0.5em minus 0.4em\relax
  Springer, 2016, pp. 88--99.

\bibitem{CPS-Datasets}
``{iTrust Labs: Datasets},''
  {\url{https://itrust.sutd.edu.sg/itrust-labs_datasets/}}, 2023, accessed:
  February 2023.

\bibitem{Tretmans96a}
J.~Tretmans, ``Test generation with inputs, outputs, and quiescence,'' in
  \emph{Proc.\ International Workshop on Tools and Algorithms for Construction
  and Analysis of Systems (TACAS 1996)}, ser. LNCS, vol. 1055.\hskip 1em plus
  0.5em minus 0.4em\relax Springer, 1996, pp. 127--146.

\bibitem{Mathur-Tippenhauer16a}
A.~P. Mathur and N.~O. Tippenhauer, ``{SWaT}: a water treatment testbed for
  research and training on {ICS} security,'' in \emph{Proc.\ International
  Workshop on Cyber-physical Systems for Smart Water Networks (CySWater@CPSWeek
  2016)}.\hskip 1em plus 0.5em minus 0.4em\relax {IEEE} Computer Society, 2016,
  pp. 31--36.

\bibitem{SWaT-Reference}
``{Secure Water Treatment (SWaT)},''
  \url{https://itrust.sutd.edu.sg/testbeds/secure-water-treatment-swat/}, 2023,
  accessed: February 2023.

\bibitem{Supplementary-Material}
``Code for {SWaT} experiments,''
  \url{https://github.com/yuqiChen94/Causally-Different-Attacks/}, 2023.

\bibitem{Goh_et-al17a}
J.~Goh, S.~Adepu, M.~Tan, and Z.~S. Lee, ``Anomaly detection in cyber physical
  systems using recurrent neural networks,'' in \emph{Proc.\ International
  Symposium on High Assurance Systems Engineering (HASE 2017)}.\hskip 1em plus
  0.5em minus 0.4em\relax {IEEE}, 2017, pp. 140--145.

\bibitem{Goldberg89a}
D.~E. Goldberg, \emph{Genetic Algorithms in Search, Optimization and Machine
  Learning}.\hskip 1em plus 0.5em minus 0.4em\relax Addison-Wesley, 1989.

\bibitem{Beer-et_al09a}
I.~Beer, S.~Ben{-}David, H.~Chockler, A.~Orni, and R.~J. Trefler, ``Explaining
  counterexamples using causality,'' in \emph{Proc.\ International Conference
  on Computer Aided Verification (CAV 2009)}, ser. LNCS, vol. 5643.\hskip 1em
  plus 0.5em minus 0.4em\relax Springer, 2009, pp. 94--108.

\bibitem{Beer-et_al12a}
------, ``Explaining counterexamples using causality,'' \emph{Formal Methods in
  System Design}, vol.~40, no.~1, pp. 20--40, 2012.

\bibitem{Chowdhury-Johnson-Csallner17a}
S.~A. Chowdhury, T.~T. Johnson, and C.~Csallner, ``{CyFuzz}: {A} differential
  testing framework for cyber-physical systems development environments,'' in
  \emph{Proc.\ Workshop on Design, Modeling and Evaluation of Cyber Physical
  Systems (CyPhy 2016)}, ser. LNCS, vol. 10107.\hskip 1em plus 0.5em minus
  0.4em\relax Springer, 2017, pp. 46--60.

\bibitem{Shrestha-Chowdhury-Csallner20a}
S.~L. Shrestha, S.~A. Chowdhury, and C.~Csallner, ``{DeepFuzzSL}: Generating
  models with deep learning to find bugs in the {Simulink} toolchain,'' in
  \emph{Proc.\ Workshop on Testing for Deep Learning and Deep Learning for
  Testing (DeepTest 2020)}.\hskip 1em plus 0.5em minus 0.4em\relax {ACM}, 2020.

\bibitem{Vigna-et_al04a}
G.~Vigna, W.~K. Robertson, and D.~Balzarotti, ``Testing network-based intrusion
  detection signatures using mutant exploits,'' in \emph{Proc.\ {ACM}
  Conference on Computer and Communications Security (CCS 2004)}.\hskip 1em
  plus 0.5em minus 0.4em\relax {ACM}, 2004, pp. 21--30.

\bibitem{DBLP:conf/fmics/MoradiASCST20}
F.~Moradi, S.~A. Asadollah, A.~Sedaghatbaf, A.~Causevic, M.~Sirjani, and C.~L.
  Talcott, ``An actor-based approach for security analysis of cyber-physical
  systems,'' in \emph{Proc.\ International Conference on Formal Methods for
  Industrial Critical Systems (FMCIS 2020)}, ser. LNCS, vol. 12327.\hskip 1em
  plus 0.5em minus 0.4em\relax Springer, 2020, pp. 130--147.

\bibitem{9141597}
R.~{Zhang}, Z.~{Cao}, and K.~{Wu}, ``Tracing and detection of {ICS} anomalies
  based on causality mutations,'' in \emph{Proc.\ Information Technology and
  Mechatronics Engineering Conference (ITOEC 2020)}, 2020, pp. 511--517.

\bibitem{DBLP:conf/ecai/IbrahimKZKP20}
A.~Ibrahim, T.~Klesel, E.~Zibaei, S.~Kacianka, and A.~Pretschner, ``Actual
  causality canvas: {A} general framework for explanation-based socio-technical
  constructs,'' in \emph{Proc.\ European Conference on Artificial Intelligence
  ({ECAI} 2020)}, ser. Frontiers in Artificial Intelligence and Applications,
  vol. 325.\hskip 1em plus 0.5em minus 0.4em\relax {IOS} Press, 2020, pp.
  2978--2985.

\bibitem{DBLP:journals/cacm/Pearl19}
J.~Pearl, ``The seven tools of causal inference, with reflections on machine
  learning,'' \emph{Communications of the {ACM}}, vol.~62, no.~3, pp. 54--60,
  2019.

\bibitem{DBLP:conf/icml/ForneyPB17}
A.~Forney, J.~Pearl, and E.~Bareinboim, ``Counterfactual data-fusion for online
  reinforcement learners,'' in \emph{Proc.\ International Conference on Machine
  Learning (ICML 2017)}, ser. Proceedings of Machine Learning Research,
  vol.~70.\hskip 1em plus 0.5em minus 0.4em\relax {PMLR}, 2017, pp. 1156--1164.

\bibitem{Annpureddy-et_al11a}
Y.~Annpureddy, C.~Liu, G.~E. Fainekos, and S.~Sankaranarayanan, ``{S-TaLiRo}:
  {A} tool for temporal logic falsification for hybrid systems,'' in
  \emph{Proc.\ International Conference on Tools and Algorithms for the
  Construction and Analysis of Systems ({TACAS} 2011)}, ser. LNCS, vol.
  6605.\hskip 1em plus 0.5em minus 0.4em\relax Springer, 2011, pp. 254--257.

\bibitem{Yamagata-et_al20a}
Y.~Yamagata, S.~Liu, T.~Akazaki, Y.~Duan, and J.~Hao, ``Falsification of
  cyber-physical systems using deep reinforcement learning,'' \emph{{IEEE}
  Transactions on Software Engineering}, vol.~47, no.~12, pp. 2823--2840, 2021.

\bibitem{Zalewski}
M.~Zalewski, ``{American fuzzy lop},'' \url{http://lcamtuf.coredump.cx/afl/},
  2017, accessed: February 2023.

\bibitem{Cha-Woo-Brumley15a}
S.~K. Cha, M.~Woo, and D.~Brumley, ``Program-adaptive mutational fuzzing,'' in
  \emph{Proc.\ {IEEE} Symposium on Security and Privacy (S{\&}P 2015)}.\hskip
  1em plus 0.5em minus 0.4em\relax {IEEE} Computer Society, 2015, pp. 725--741.

\bibitem{Boehme-et_al17a}
M.~B{\"{o}}hme, V.~Pham, M.~Nguyen, and A.~Roychoudhury, ``Directed greybox
  fuzzing,'' in \emph{Proc.\ {SIGSAC} Conference on Computer and Communications
  Security (CCS 2017)}.\hskip 1em plus 0.5em minus 0.4em\relax {ACM}, 2017, pp.
  2329--2344.

\bibitem{DBLP:conf/uss/Chen0MLWZJS19}
Y.~Chen, Y.~Jiang, F.~Ma, J.~Liang, M.~Wang, C.~Zhou, X.~Jiao, and Z.~Su,
  ``Enfuzz: Ensemble fuzzing with seed synchronization among diverse fuzzers,''
  in \emph{Proc.\ {USENIX} Security Symposium ({USENIX} Security 2019)}.\hskip
  1em plus 0.5em minus 0.4em\relax {USENIX} Association, 2019, pp. 1967--1983.

\bibitem{DBLP:conf/icse/NguyenP0L020}
T.~D. Nguyen, L.~H. Pham, J.~Sun, Y.~Lin, and Q.~T. Minh, ``{sFuzz}: an
  efficient adaptive fuzzer for solidity smart contracts,'' in \emph{Proc.\
  International Conference on Software Engineering (ICSE 2020)}.\hskip 1em plus
  0.5em minus 0.4em\relax {ACM}, 2020, pp. 778--788.

\bibitem{Godefroid-Kiezun-Levin08a}
P.~Godefroid, A.~Kiezun, and M.~Y. Levin, ``Grammar-based whitebox fuzzing,''
  in \emph{Proc.\ ACM SIGPLAN Conference on Programming Language Design and
  Implementation (PLDI 2008)}.\hskip 1em plus 0.5em minus 0.4em\relax {ACM},
  2008, pp. 206--215.

\bibitem{Zeller-et_al19a}
\BIBentryALTinterwordspacing
A.~Zeller, R.~Gopinath, M.~B{\"o}hme, G.~Fraser, and C.~Holler, ``The fuzzing
  book.''\hskip 1em plus 0.5em minus 0.4em\relax CISPA Helmholtz Center for
  Information Security, 2023, accessed: February 2023. [Online]. Available:
  \url{https://www.fuzzingbook.org/}
\BIBentrySTDinterwordspacing

\bibitem{Johnson-Brun-Meliou20a}
B.~Johnson, Y.~Brun, and A.~Meliou, ``Causal testing: understanding defects'
  root causes,'' in \emph{Proc.\ International Conference on Software
  Engineering (ICSE 2020)}.\hskip 1em plus 0.5em minus 0.4em\relax {ACM}, 2020,
  pp. 87--99.

\bibitem{Ruscito}
A.~Ruscito, ``pycomm,'' \url{https://github.com/ruscito/pycomm}, 2019,
  accessed: February 2023.

\end{thebibliography}

%\newpage
\appendix

\section*{Equivalence Class Constructions}

\begin{proposition}[Strong capability-set equivalence]\label{prop:strong-cap-set-equiv}
	\emph{Let $t$ denote a successful test on CPS $\mathcal{P}$. A strategy $\mathrm{Excl}([t]_{\simeq})$ can be constructed such that for every successful test $t'$ on $\mathcal{P}$:}
	\[ \pi_{t'} \in L(\mathrm{Excl}([t]_{\simeq})) \ \ \text{\emph{if and only if}}\ \ t' \not\in [t]_\simeq \]

    \noindent Construction. \emph{Suppose that $\mathrm{CSet}(\pi_t) = \{y_1, \dots, y_n\}$.
    Let us define $Y^\ast = 2^{\mathrm{CSet}(\pi_t)}\setminus\mathrm{CSet}(\pi_t)$. Then,}
\begin{equation*}
	\begin{split}
		\mathrm{Excl}&([t]_{\simeq}) =\\
		& \hspace{0.05in}(\{q_\ast\} \cup \{q_Y\mid Y\in Y^\ast\},\\
		& \hspace{0.05in} \{(q_\ast, (\mathrm{true}, \_, \mathrm{true}), q_\ast)\}\vspace{-0.5in} \\%[-8pt]
		&\ \ \ \bigcup\limits_{Y\in Y^\ast} \hspace{-0.00in}\bigl\{ (q_Y, (\mathrm{true}, \_\setminus\mathrm{CSet}(\pi_t) \neq\ \emptyset ), q_\ast) \bigr\} \\
		&\ \ \ \bigcup\limits_{\hspace{0.2in}\substack{Y,Y'\in Y^\ast\\. Y\subseteq Y'}} \hspace{-0.15in}\bigl\{ (q_Y, (\mathrm{true}, Y'\setminus Y \subseteq \_ \wedge \_ \subseteq Y'), q_{Y'}) \bigr\},\\
		& \hspace{0.05in} q_\emptyset)
	\end{split}
\end{equation*}

\noindent \emph{where $q_\ast$ and each $q_Y$ are fresh strategy states.}
\qed
\end{proposition}

\begin{figure*}[!t]
    \centering
    \subfloat{{\includegraphics[width=0.45\linewidth]{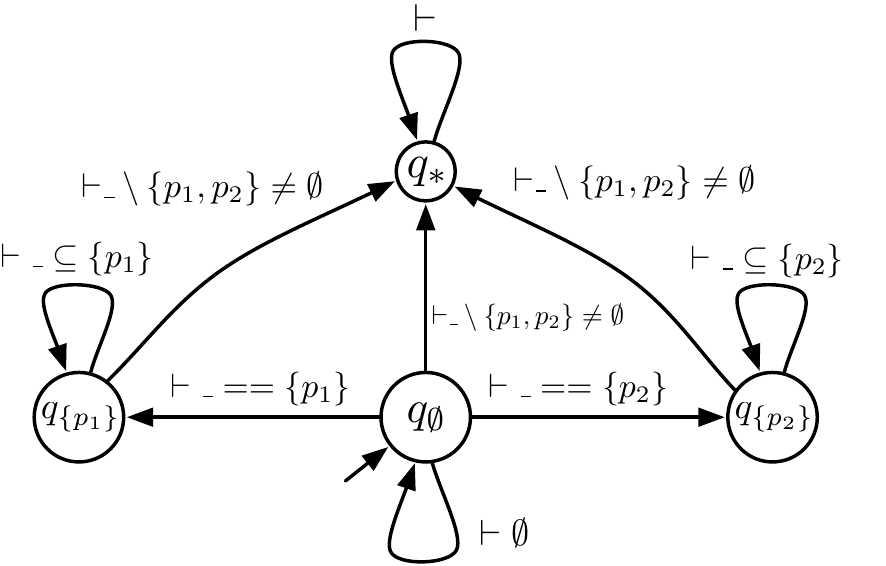} }}%
    \qquad
    \subfloat{{\includegraphics[width=0.45\linewidth]{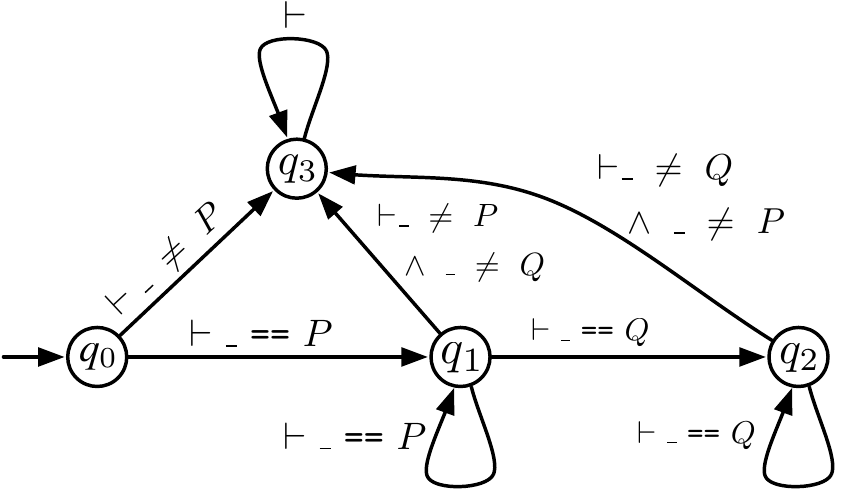} }}%
    \caption{\emph{Left:} constructing $\mathrm{Excl}([t]_{\simeq})$ for $\mathrm{CSet}(\pi_t) = \{p_1, p_2\}$; \emph{Right:} $\mathrm{Excl}([t]_\cordequivsub)$ for $\mathrm{COrd}(\pi_t) = PQP$}%
    \label{fig:construction_strong_capset}
    \label{fig:construction_strong_capord}%
\end{figure*}

For example, suppose that $\pi_t = PPPQQQPPP$ is the history of a successful test, where $P = \{p_1\}$, $Q = \{p_1,p_2\}$, and thus $\mathrm{CSet}(\pi_t) = \{p_1, p_2\}$. Then $\mathrm{Excl}([t]_{\simeq})$ is as depicted in Figure~\ref{fig:construction_strong_capset}. Note that the constraints of some transitions have been simplified. Intuitively, the strategy characterises tests that do not involve \emph{exactly} the same set of capabilities as $t$. This can be achieved by using strictly more capabilities than $t$, or by using strictly fewer capabilities than $t$. In particular, the strategy excludes any tests achieved by using \emph{exactly} the capabilities $p_1$ and $p_2$ (whether together or in different transitions) and no others.

\begin{proposition}[Strong capability-order equivalence]\label{prop:strong_capability-order_equivalence}
	\emph{Let $t$ be a successful test on CPS $\mathcal{P}$. A strategy $\mathrm{Excl}([t]_\cordequivsub)$ can be constructed such that for every successful test \mbox{$t'$ on $\mathcal{P}$:}}
	\[ \pi_{t'} \in L(\mathrm{Excl}([t]_\cordequivsub)) \ \ \text{\emph{if and only if}}\ \ t' \not\in [t]_\cordequivsub \]

    \noindent Construction. \emph{Suppose that $\mathrm{COrd}(\pi_t) = Y_1Y_2\dots Y_k$. Define:}
\begin{equation*}
	\begin{split}
		\mathrm{Excl}&([t]_\cordequivsub) =\\
		&(\{q_0, q_1, \dots, q_k\}, \\
		& \{ (q_0, (\mathrm{true}, \_ \neq\ Y_1), q_k), (q_k, (\mathrm{true}, \mathrm{true}), q_k) \} \\
		&\ \ \ \bigcup\limits_{i=1}^{k-1} \bigl\{ (q_{i-1}, (\mathrm{true}, \_\ \mt{==}\ Y_i), q_i), (q_i, (\mathrm{true}, \_\ \mt{==}\ Y_i), q_i),  \\
		&\ \ \ \hspace{0.25in} (q_i, (\mathrm{true},  \_\ \neq\ Y_i \wedge \_\ \neq\ Y_{i+1}), q_k) \bigr\},\\
		& q_0 )
	\end{split}
\end{equation*}

\noindent \emph{where $q_0, q_1, \dots, q_k$ are fresh strategy states.}
\qed

\end{proposition}

For example, suppose that $\pi_t = PPPQQQPPP$ is the history of a successful test, where $P = \{p_1\}$, $Q = \{p_1,p_2\}$, and $\mathrm{COrd}(\pi_a) = PQP$. Then $\mathrm{Excl}([t]_\cordequivsub)$ is as depicted in Figure~\ref{fig:construction_strong_capord}. The strategy characterises tests that do \emph{not} consist of usages of $P$, followed by usages of $Q$, followed by usages of $P$ again. Intuitively, it achieves this by using states $q_1$ and $q_2$ to `track' the usages of $P$s followed by $Qs$, with transitions to $q_3$ whenever a capability is used that makes it impossible to achieve the $P..Q..P..$~pattern. Note that from $q_2$ there are no transitions supporting the usage of $P$.

\section*{Proof of Theorem~\ref{thm:composition}}

\begin{proof}[Proof Sketch]
    First, it is required to show that a capability history $\pi_{t'} = Y_1Y_2Y_3\cdots$ is in $L(\mathcal{T}\parallel\mathrm{Excl}([t]_{\sim}))$ if and only if $\pi_{t'} \in L(\mathcal{T})$ and $\pi_{t'} \in L(\mathrm{Excl}([t]_{\sim}))$.
    This can be obtained by showing that any test trace in $\mathcal{T}\parallel\mathrm{Excl}([t]_{\sim})$ that derives $\pi_{t'}$ can be mapped to a corresponding trace of states and transitions in $\mathcal{T}$ and $\mathrm{Excl}([t]_{\sim})$ (and vice versa).
    This is evident from the definition of $\parallel$ as the composed strategy explicitly tracks (disjoint) state identifiers, combines all possible pairs of transitions, and fully synchronises on the steps of both strategies.
    The result is then obtained by appealing to the relevant Proposition of the equivalence class defined by the user (e.g.~Propositions~\ref{prop:capability-set_equivalence}, \ref{prop:strong-cap-set-equiv}, and~\ref{prop:strong_capability-order_equivalence}), i.e.~that $\pi_{t'} \in L(\mathrm{Excl}([t]_{\sim}))$ if and only if $t' \not\in [t]_\sim$.
\end{proof}

\section*{Implementation for SWaT}
\label{sec:implementing_for_swat}

The formalisation, equivalence classes, and algorithms we have presented are completely general: they can be applied to any test scenario characterised as performing sequences of sensor/actuator manipulations, and they do not require any modelling of the targeted CPS (which would be difficult to do in general, owing to the presence of physical processes). Our theory can be applied to a real system by identifying the sensors/actuators a tester can manipulate (along with their possible readings/configurations), an equivalence class of interest, and a test strategy of interest. Furthermore, in order to guide the search for effective tests from a strategy, it requires some black box predictive model of the system's behaviour (e.g.~a simulator, or a neural network trained on sensor/actuator data of the real system).

To demonstrate the effectiveness of our theory in practice, we implemented causal fuzzing for SWaT as a suite of Python scripts, which are available online~\cite{Supplementary-Material}. Our implementation is able to handle strong capability-set equivalence, strong capability-order equivalence, most importantly, $Y$-set equivalence, where $Y$ is taken to be the \emph{causal} capabilities of an identified test.

In Algorithm~\ref{alg:derive_attack}, to constrain the search space, we limit walks to three transitions, but use a longer fixed time interval of 10 minutes (determined experimentally; Section~\ref{sec:evaluation}). As our prediction model, we make use of a high-fidelity simulator for SWaT~\cite{Supplementary-Material}, but note that other kinds of prediction models based on machine learning are available too~\cite{Goh_et-al17a,Chen-Poskitt-et_al19a,Chen-Xuan-Poskitt-et_al20a}. We define simple objective functions in terms of the known safe operational ranges of sensors (similar to the fuzzer of~\cite{Chen-Poskitt-et_al19a}). To test whether a manipulation is causal (Algorithm~\ref{alg:causal_events}), we make use of the actual testbed for all tests except those that target tank over- or underflows. For those tests, we use the simulator to avoid resource wastage, testing only the final output $t_\mathrm{min}$ on the real system.

Once a potential test has been identified and pruned, the rest of Algorithm~\ref{alg:overall_algorithm} (i.e.~identifying whether the potential test is \emph{actually} successful) is executed on the SWaT testbed itself. To assess the effects of utilising capabilities, we fuzz the sensor readings and actuator commands accordingly over the network using the pycomm~\cite{Ruscito} package.

\section*{Causally Different Tests Found}\label{sec:attack_table}

Table~\ref{tab:causality_sets} lists the causality sets identified in our evaluation (Section~\ref{sec:evaluation}; Experiment \#2).

\begin{table*}[!t]
\caption{Causal capability sets in tests identified by our fuzzer}\label{tab:causality_sets}
\centering\small
\begin{tabular}{|l||l|}
\hline
\textbf{Test Goal}        & \textbf{Causal Capability Sets}                                                                                                                                                                                                                                                                                                                         \\ 
\hline\hline

FIT101 (High)  & \{{[}MV101, open]\} \\
\hline
FIT201 (High)  & \{{[}MV201, open], [P101, on], [P102, on]\}    \\
\hline
FIT601 (High)  & \{{[}MV301, open], [MV303, open], [MV502, open],\\
& \hspace{10pt} [P602, on]\}    \\
\hline

FIT101 (Low)  & \{{[}MV101, close]\}                                                                                                                                                                                                                                                                                                                                                \\ 
\hline
FIT201 (Low)  & \{{[}MV201, close]\}\\
    & \{{[}P101, off], [P102, off]\}                                                                                                                                                                                                                                                                                                            \\ 
\hline
FIT301 (Low)  & \{{[}MV302, close]\}\\
    & \{{[}P301, off], [P302, off]\}                                                                                                                                                                                                                                                                                                            \\ 
\hline
FIT401 (Low)  & \{{[}MV501, close]\}                                                                                                                                                                                                                                                                                                                                               \\ 
\hline
FIT501 (Low)  & \{{[}MV501, close], [MV503, close], [MV504, close]\}                                                                                                                                                                                                                                                                                                                \\ 
\hline
FIT601 (Low)  & \{{[}P602, off]\}                                                                                                                                                                                                                                                                                                                                                 \\ 
\hline
DPIT301 (Low) & \{{[}MV301, open], [MV303, open], [P602, on]\}\\
    & \{{[}MV302, close]\}\\
    & \{{[}P301, off], [P302, off]\}                                                                                                                                                                                                                \\ 
\hline

LIT101 (High)  & \{{[}MV101, open]\} \\
    & \{{[}MV101, open], [P101, off], [P102, off]\} \\
    & \{{[}MV101, open], [MV201, close]\}   \\
\hline

LIT301 (High)  & \{{[}MV201, open], [P101, on]\} \\
    & \{{[}MV201, open], [P102, on]\} \\
    & \{{[}MV201, open], [P101, on], [P301, off], [P302, off]\} \\ 
    & \{{[}MV201, open], [P102, on], [P301, off], [P302, off]\} \\
    & \{{[}MV201, open], [P101, on], [MV302, close]\} \\
    & \{{[}MV201, open], [P102, on], [MV302, close]\} \\
\hline

LIT401 (High)  & \{{[}MV302, open], [P301, on]\} \\
    & \{{[}MV302, open], [P302, on]\} \\
\hline

LIT101 (Low)  & \{{[}MV101, close], [MV201, open], [P101, on]\}\\
    & \{{[}MV101, close], [MV201, open], [P102, on]\}                                                                                                                                                                                                                     \\ 
\hline
LIT301 (Low)  & \{{[}MV201, close], [MV302, open], [P301, on]\}\\
    & \{{[}MV201, close], [MV302, open], [P302, on]\}                                                                                                                                                                                                                     \\ 
\hline
LIT401 (Low)  & \{{[}MV302, close], [P401, on]\} \\
    & \{{[}P301, off], [P302, off], [P401, on]\} \\
    & \{{[}MV302, close], [P402, on]\}\\
	& \{{[}P301, off], [P302, off], [P402, open]\}                               \\
	\hline
\end{tabular}
\end{table*}

\end{document}